\documentclass[
hidelinks,onefignum,onetabnum]{siamart250211}

\usepackage{lipsum}
\usepackage{amsfonts}
\usepackage{graphicx}
\usepackage{epstopdf}
\usepackage{algorithmic}
\ifpdf
  \DeclareGraphicsExtensions{.eps,.pdf,.png,.jpg}
\else
  \DeclareGraphicsExtensions{.eps}
\fi

\newsiamremark{remark}{Remark}
\newsiamremark{hypothesis}{Hypothesis}
\crefname{hypothesis}{Hypothesis}{Hypotheses}
\newsiamthm{claim}{Claim}
\newsiamremark{fact}{Fact}
\crefname{fact}{Fact}{Facts}

\headers{Reinforced-GANs for Market Equilibrium Models}{A. Kratsios, X. Shi, Q. Sun, Z. Zhang}

\title{Generative Market Equilibrium Models with\\ 
Stable Adversarial Learning via Reinforcement Link\thanks{Written  on April 4th, 2025.
\funding{A.\ Kratsios acknowledge financial support from NSERC Discovery Grant No.\ RGPIN-2023-04482 and No.\ DGECR-2023-00230. X.\ Shi acknowledge financial support from NSERC Discovery Grant No. RGPIN-2024-04569 and No. DGECR-2024-004.
Q.\ Sun acknowledge financial support from an NSERC Discovery Grant No. RGPIN-2018-06484.
They also acknowledge that resources used in preparing this research were provided, in part, by the Province of Ontario, the Government of Canada through CIFAR, and companies sponsoring the Vector Institute (\href{https://vectorinstitute.ai/partnerships/current-partners/}{https://vectorinstitute.ai/partnerships/current-partners/}). The authors gratefully acknowledge constructive comments and insightful discussions with 
Paolo Guasoni, Jiequn Han, Sebastian Jaimungal, Johannes Muhle-Karbe, and Martin Larsson
}
}
}

\author{
Anastasis Kratsios\thanks{{Vector Institute}, and McMaster University, Department of Mathematics; \texttt{kratsioa@mcmaster.ca}}, 
\and Xiaofei Shi\thanks{University of Toronto, Department of Statistical Sciences; \texttt{xf.shi@utoronto.ca}}, 
\and Qiang Sun\thanks{University of Toronto, Department of Statistical Sciences; \texttt{qiang.sun@utoronto.ca}.}, 
\and Zhanhao Zhang\thanks{Cornell University, Operations Research and Information Engineering; \texttt{zz564@cornell.edu}.}
}

\usepackage{amsopn}

\ifpdf
\hypersetup{
  pdftitle={Generative Market Equilibrium Models with 
Stable Adversarial Learning via Reinforcement Link},
  pdfauthor={A. Kratsios, X. Shi, Q. Sun, Z. Zhang}
}
\fi

\usepackage{hyperref} 
    \hypersetup{
    	colorlinks = true,
    	linkcolor = blue,
    	anchorcolor = blue,
    	citecolor = blue,
    	filecolor = blue,
    	urlcolor = blue
    }

\usepackage{mathrsfs} 
\newcommand{\mfT}{{\mathfrak{T}}}
\newcommand{\mfN}{{\mathfrak{N}}}
\newcommand{\mfS}{{\mathfrak{S}}}
\newcommand{\mfY}{{\mathfrak{Y}}}
\newcommand{\scrF}{\mathscr{F}}
\newcommand{\PP}{\mathbb{P}}
\newcommand{\EE}{\mathbb{E}}
\newcommand{\RR}{\mathbb{R}}
\newcommand{\cA}{\mathcal{A}}
\newcommand{\cF}{\mathcal{F}}
\newcommand{\cH}{\mathcal{H}}
\newcommand{\cN}{\mathcal{N}}
\newcommand{\tr}{\mathop{\mathrm{tr}}}
\newcommand{\sign}{\mathop{\mathrm{sign}}}
\newcommand{\eqdef}{\ensuremath{\stackrel{\mbox{\upshape\tiny def.}}{=}}}

\newcommand{\XS}[1]{{\color{violet} #1}}

\begin{document}

\maketitle

\begin{abstract}
We present a general computational framework for solving continuous-time financial market equilibria under minimal modeling assumptions while incorporating realistic financial frictions, such as trading costs, and supporting multiple interacting agents. Inspired by generative adversarial networks (GANs), our approach employs a novel generative deep reinforcement learning framework with a decoupling feedback system embedded in the adversarial training loop, which we term as the \emph{reinforcement link}. This architecture stabilizes the training dynamics by incorporating feedback from the discriminator. Our theoretically guided feedback mechanism enables the decoupling of the equilibrium system, overcoming challenges that hinder conventional numerical algorithms. Experimentally, our algorithm not only learns but also provides testable predictions on how asset returns and volatilities emerge from the endogenous trading behavior of market participants, where traditional analytical methods fall short. The design of our model is further supported by an approximation guarantee.
\end{abstract}



\section{Introduction}
Equilibrium models are highly valued in financial markets as they provide a framework for understanding how asset prices and other financial variables are determined through the endogenous trading behaviors of market participants. In particular, even the most frequently traded assets have limited liquidity provided in the market. Hence, the dynamic interplay between asset prices and agents' trading behaviors under the presence of trading costs has been a focal point of extensive research; see~\cite{adam.al.15,buss2019dynamic,heaton1996evaluating, lo.al.04}. 
To establish a theoretical foundation for the impact of illiquidity, it is essential to formulate equilibrium asset pricing models. In these models, price levels, returns, and volatilities are not treated as exogenous inputs, but instead emerge endogenously through the matching of supply and demand. This equilibrium approach enables a deeper understanding of how price characteristics are influenced by market liquidity.

Analyzing equilibrium models with trading costs is notoriously challenging, as limited liquidity and equilibrium asset pricing are complex issues. These difficulties are compounded when asset price dynamics are determined endogenously in the presence of trading frictions, which significantly complicates the agents' individual optimization problems. In addition, representative agents cannot capture the impact of trading costs as they do not account for trades between individual market participants. Even in tractable models, such as the Linear-Quadratic (LQ) framework \cite{gonon2021asset, muhle2023equilibrium, shi2020equilibrium, noh2022price}, the individual optimization problem remains nontrivial. Recent work has focused on models with random fluctuations in asset prices and trading volume, analyzing quadratic costs on trading rates \cite{almgren.03, muhle2023equilibrium, shi2020equilibrium}. However, empirical estimates of trading costs typically follow a power law with an exponent around 3/2, see~\cite{almgren.al.05, lillo.al.03}. The excess equilibrium return $\mu$ can be derived from market-clearing conditions in two-agent markets with nonlinear costs. However, the resulting fully coupled forward-backward stochastic differential equations (FBSDEs) fall beyond the scope of known well-posedness results. In markets with more than two agents, $\mu$ is only implicitly defined, making even advanced deep learning-based numerical methods, such as the FBSDE Solver from \cite{han2018solving}, inapplicable. Although there are tailored numerical methods for specific incomplete financial equilibria in discrete-time \cite{buss2019dynamic, dumas2012incomplete}, a general framework for continuous-time equilibrium models remains elusive.

\begin{figure}[htp!]
    \centering
        \includegraphics[width=.35\textwidth]{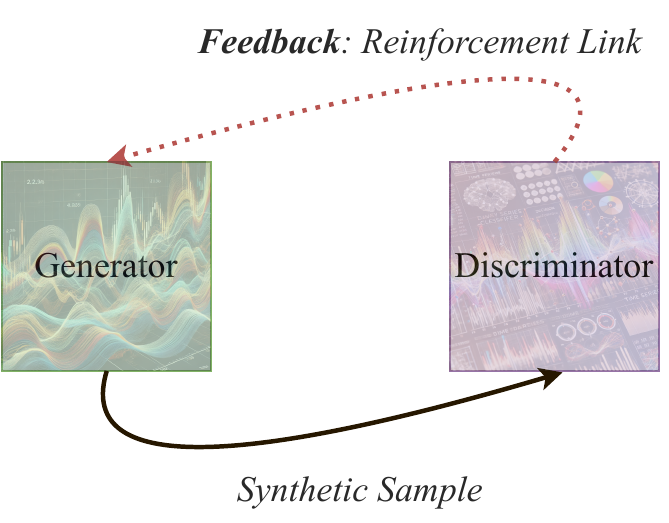}
        \caption{\textbf{Training Pipeline - The Reinforement Link}: Standard adversarial training (\textit{bottom arrow only}) involves passing samples from the generator (our model) to the discriminator (effectively our loss function), which determines whether a sample is synthetic or real. Our training pipeline (\textit{both top and bottom arrow}) stabilizes this inherently unstable process by incorporating a \textit{feedback} mechanism, our so-called ``reinforcement link'', allowing the generator to leverage the discriminatory decisions when iteratively refining its sampling strategy.  }
        \label{fig:GAN}
    \end{figure}

\paragraph{Contributions}
We propose a modern deep learning approach to overcome the limitations of classical analytic and traditional computational approaches to understanding market equilibria.  Our approach leverages the power of generative models over the spaces of trading strategies and square-integrable martingales.  
Illustrated by Figure~\ref{fig:GAN}, the training of our generative models build on the generative \textit{adversarial} networks (GANs)~\cite{goodfellow2014generative}, methodology where we stabilize the training procedure by allowing the generator (our model) to receive information from the discriminator during training, which we term~\textit{reinforcement link}.  
The effect of our training and theoretically founded AI-powered equilibrium model is reflected across our numerical experiments.  Our technology allows us to compute market equilibria in previously, both analytically and computationally, intractable and realistic multi-agent settings.  Further, it is more accurate than the available computationally manageable first-order approximations, derived under additional stylized assumptions in~\cite{shi2020equilibrium}.
Lastly, as a sanity check, we verify that our proposed method recovers classical analytic results in the LQ preference case.

From an approximation perspective, our randomized time-horizon technique for deep learning approximations to stochastic processes offers a novel way to embed the complexity of the neural network within the structure of a small (positive) time horizon, avoiding dimensional dependence on network depth and width. This coupling method is a temporal analogue of the technique in \cite{kratsios2025approximation}, which delegates complexity across multiple "expert models" locally in space.  The key point is that, by embedding complexity into the stopping time, we may avoid requiring more structure or regularity of the target function, e.g.\ in~\cite{barron1993universal,mhaskar2016deep} to achieve efficient approximation rates, e.g. in~\cite{petersen2018optimal,gonon2023approximation,neufeld2023universal}, of the target function in order to avoid the cursed \textit{min-max optimal} worst-cast approximation rates in deep learning~\cite{yarotsky2017error,kratsios2022universal,petersen2018optimal}. 

In addition, we find a light universal representation of a broad range of continuous-time financial markets (Theorem~\ref{thrm:AprxBaby}). Depending on parameters with numbers linear in the reciprocal approximation error,  our approximation result guarantees that a broad class of \textit{light} controlled neural SDEs can approximate the target-controlled SDE in a pathwise sense, with high probability.  This differs from the available approximation guarantees for (non-controlled but possibly with jumps) neural SDEs, e.g.~\cite{gonon2023deep,biagini2024approximation}, which also require a super-linear polynomial number of parameters.   Our key approximation-theoretic insight is the use of small randomized time horizons guaranteeing, which ensures that all process paths are highly localized with high probability on the relevant random time interval.


Our code pipeline and implementation details are accessible via the following link:
\href{https://github.com/xf-shi/Reinforced-GAN}{https://github.com/xf-shi/Reinforced-GAN}.

\paragraph{Organization of The Paper}
Our paper is organized as follows. Section~\ref{s:Prelims} covers all necessary preliminaries, from notation to the markets we consider.
Section~\ref{sec:algo}, we introduce our carefully designed Reinforced-GAN algorithm to overcome these difficulties and provide numerical examples to illustrate the power of our methods in Section~\ref{sec:exp}. 
All proofs are relegated to our appendices.

\paragraph{Notation}
We fix a filtered probability space $(\Omega,\scrF,(\scrF_t)_{t \in \mfT},\PP)$ with finite time horizon $\mfT = [0,T]$, where the filtration is generated by a $d$-dimensional standard Brownian motion $B=(B_t)_{t \in \mfT}$. Throughout, let $\|\cdot\|$ be the 2-norm of a real-valued vector.


\section{Market Equilibrium Models}
\label{s:Prelims}
\subsection{Risk Sharing Economy}
We consider a financial market with $m+1$ assets. The first one is safe and earns a constant interest rate $r>0$ \footnote{Our framework can be extended to equilibrium without frictions or even when the interest rate is determined endogenously through the consumption market clearing conditions.}. The other $m$ assets are risky with (cum-dividend) price  dynamics
\begin{align}\label{eq:Sdyn}
dS_t=(rS_t +\mu_t) dt+\sigma_t dB_t, \quad S_T=\mfS.
\end{align}
Here, the $\scrF_T$-measurable liquidating dividend $\mfS$ is given exogenously. In contrast, the $\RR^d$-valued expected excess returns $\mu$ and the $\RR^{m\times d}$-valued volatility process $\sigma$ are to be determined \emph{endogenously} by matching the agents' demand to the fixed supply $ s \geq 0$ of the risky asset. Together with the initial stock price, participating agents can observe the expected excess returns $\mu$ and the volatility $\sigma$, but do not have access to see others' positions or trading strategies. We use $(S_0, \mu,\sigma)$
to denote the public information in the market. 

We consider an economy with agents indexed by $n\in\mfN = \{1,2, \ldots,N\}$. Each of these agents receives a  random endowment stream $\zeta_{n,t}$ with the following dynamics:
\begin{align}\label{eq:endowment}
d\zeta_{n,t} =b_{n,t}dt + \xi_{n,t} dB_t. 
\end{align}
Here, the drift $b_n$ and the volatility $\xi_n$ processes of Agent-$n$'s random endowment  are also diffusion processes with known dynamics, hence can be simulated.
We assume that the safe asset is perfectly liquid, but trades of the risky assets incur
dead-weight trading costs due to shortage of liquidity. Therefore, we focus on absolutely continuous trading strategies $\dot\varphi $, where Agent-$n$ controls their trading rates on the $m$ risky assets:
\begin{align}\label{eq:trading rate}
d\varphi_{n,t} = \dot{\varphi}_{t,n}\; dt, \qquad \varphi_{n,0}\in\RR^{m}. 
\end{align}
and penalize the trading rates $\dot\varphi$ with an \emph{instantaneous} trading costs
$G(\dot\varphi_{t};\Lambda_t)$. 
Here, $G: \RR^{m} \to \RR_+$ is strictly convex and differentiable, and $G(\dot\varphi;\Lambda)$ is strictly positive for every $\dot\varphi\in\RR^m/\{0\}$. $\Lambda\in\RR^{m\times m}$ represents the liquidity parameter, which can be a symmetric, positive definite matrix, or an adaptive multi-dimensional stochastic process.
When there is only one stock in the market, i.e. $m=1$, our model nests the general power trading costs with $G(\dot\varphi; \Lambda) = \Lambda|\dot\varphi|^q/q, q\in(1,2]$, which is proposed in~\cite{almgren.03} and has been studied in market models such as~\cite{garleanu.pedersen.13,garleanu.pedersen.16, guasoni.weber.18, muhle2023equilibrium, shi2020equilibrium}. The limiting case in sending $q\downarrow 1$ corresponds to proportional trading costs, which have been studied intensely going back to \cite{constantinides.magill.76, davis.norman.90,dumas.luciano.91, shi2020equilibrium}. To tackle this setting, we can instead parametrize the individual agent's no-trade regions, and algorithm with the same spirit will follow. Moreover, our model also allows the trading costs to fluctuate randomly over time, in that the liquidity parameter $\Lambda_t$ is a stochastic process. This allows us to model the fluctuations of liquidity over time -- ``liquidity risk'' in the terminology of \cite{acharya.pedersen.05,dufresne.al.19}. 


\subsection{Individual Optimization Problems}
With an initial wealth $W_{n,0}\in\RR$, the wealth process of  Agent-$n$ corresponding to a generic trading strategy $\dot\varphi$ and consumption $c$ follows the dynamics
\begin{align}\label{eqn:wealth generic n}
dW_{n,t}^{\dot\varphi} 
&= d\zeta_{n,t} + \varphi_{n,t}^\top  dS_t +r\left(W_{n,t}^{\dot\varphi} - \varphi_{n,t}^\top S_t\right)dt - G(\dot\varphi_t; \Lambda_t) dt - c_tdt\notag\\
&=\left(rW_{n,t}^{\dot\varphi} + \varphi_{n,t}^\top \mu_t + b_{n,t} - G(\dot\varphi_t;\Lambda_t )-c_t\right)dt +\left(\varphi_{n,t}^\top \sigma_t +\xi_{n,t}\right) dB_t.
\end{align}
Agent-$n$ seeks a trading and consuming strategy to optimize their objective functional, 
while only receiving their endowments and observes the public information $(S_0,\mu_t, \sigma_t)$ adaptively.
They do not have access to other agents' strategies or consumption information. The objective functional for Agent-$n$ is 
\begin{align}\label{eqn:geneeral target}
J_n(\dot\varphi,c) 
&=\EE\left[\int_0^T  f_n(t,  {\varphi}_{n,t},W^{\dot\varphi,c}_{n,t}, \dot\varphi_t,c_t; \mu_t, \sigma_t, \Lambda_t) \;dt + g_n(\varphi_{n,T}, W^{\dot\varphi}_{n,T})\right]. 
\end{align}
We make the mild assumption that for each $n\in\mfN$ and fixed time $t\in\mfT$,
$f_n$ and $g_n$ are $C^1$ functions and strictly concave in wealth $W_{n,t}^{\dot\varphi,c}$, current stock position $\varphi_{n,t}$, trading rates $\dot\varphi_t$, and consumption $c_t$, and $\partial f_n(t,\varphi, w, \dot\varphi, c;\mu_t, \sigma_t, \Lambda_t,\gamma_n)/\partial w\geq0$, $\partial g_n(\varphi, w;\gamma_n)/\partial w\geq0$ for all $w\in\RR$. To ensure that the expectation of the objective functional~\eqref{eqn:geneeral target} is well-defined, the trading rates $\dot\varphi$ and consumption $c$ need to satisfy certain integrability condition, and we use $\cA$ to denote the admissible strategy set for all $(\dot\varphi,c)$. 

Our setting includes the tractable LQ model with general nonlinear trading costs, the exponential utility, and economically solid logarithmic and power utilities. 
\begin{example}[Linear-Quadratic (LQ) Preference]
\label{ex:LQ}
We recover the \textit{LQ preference} via
\begin{align}\label{target:LQ}
J_n(\dot\varphi) = \EE\left[\int_0^T \varphi_t^\top\mu_t  - \frac{\gamma_n}{2}\left\|\varphi_t^\top\sigma_t+\xi_{n,t}\right\|^2 - G(\dot\varphi_t;\Lambda_t)\; dt\right].
\end{align}
Notice that for LQ preference, researchers usually do not take consumptions into consideration; hence, we omit $c$ in the preference $J_n$ for each agent. 
\end{example}
\begin{example}[Exponential Utility]
\label{ex:ExpUt}
With $\gamma_n>0$ as the risk aversion for Agent-$n$, the classical \textit{exponential utility} is recovered by setting 
\begin{align}\label{target:exp}
J_n(\dot\varphi,c)
= \EE\Biggl[-\int_0^T\frac{1}{\gamma_c\gamma_n}\exp(-\gamma_c\gamma_n c_t)dt -\frac{1}{\gamma_n} \exp(-\gamma_n W_{n,T}^{\dot\varphi,c})\Biggr].
\end{align}
\end{example}
\begin{example}[Power Utility]
\label{ex:PowerUtility}
With $\gamma_n>0$ as the risk aversion for Agent-$n$, the economically solid \textit{power utility} is obtained by
\begin{align}\label{target:power}
J_n(\dot\varphi,c)
= \EE\Biggl[
\frac{\left(W_{n,T}^{\dot\varphi,c}\right)^{1-\gamma_n}-1}{1-\gamma_n}\Biggr]
.
\end{align}
Note that, upon setting, $\gamma_n=1$, one recovers the \textit{logarithmic utility}. 
\end{example}


\subsection{Equilibrium}

\begin{definition}
\label{frictional equilibrium}
Suppose the agents' initial positions satisfy
$\sum_{n\in\mfN}\varphi_{n,0}=s$.
A price process $S$ following~\eqref{eq:Sdyn} for the risky assets is a \emph{Radner equilibrium} with trading costs if:
\begin{enumerate}
\item[i)]
(\emph{Individual Optimality}) 
the individual optimization problem~\eqref{eqn:geneeral target} has a solution $(\dot\varphi_{n}, c_n)\in\cA$  for each agent $n\in\mfN$;
\item[ii)]
(\emph{Market Clearing}) 
the agents' total demand matches the supply of the risky assets at all times, in that $\sum_{n\in\mfN}\dot\varphi_{n,t}=0$ for all $t \in [0,T]$.
\end{enumerate}
\end{definition}





Under general nonlinear trading costs, both individual optimization and the corresponding equilibria become significantly more involved.  When the number of agents is limited to two, the solution of equilibrium models can be characterized by systems of fully-coupled nonlinear forward-backward stochastic differential equations (FBSDEs), see~\cite{gonon2021asset}. These FBSDEs usually fail outside of the well-posedness literature. Moreover, the current machine learning based numerical algorithms, such as~\cite{han2018solving}, only work when the time horizon is not too long. 
With more than two agents, the excess return $\mu$, which is the generator for the BSDEs~\eqref{eq:Sdyn}, can only be expressed in an implicit form, which fails outside of known literature. Indeed, the aforementioned approaches fail due to this implicit form issue.

\section{Reinforced-GANs for Equilibrium Models}\label{sec:algo}
One key observation is that, with the equilibrium asset prices dynamics, each agent's individual optimization problem \emph{decouples} from the fully-coupled system. 

A key idea is to \textit{separate} the learning task into two components: (a) deriving each agent's optimal trading policy for a generic price dynamic, and (b) determining the public information $(S_0, \mu_t, \sigma_t)_{t\ge 0}$ for the price process~\eqref{eq:Sdyn} that ensures market clearing and satisfies the terminal liquidation condition.
This separation naturally connects to the GAN framework, one of the most widely used deep learning architectures. Traditional GANs pass learned information from the generator to the discriminator, but the discriminator's learned results do not feed back into the generator network. To adapt the GAN framework for numerical algorithms in financial equilibrium models, we introduce a reinforced link that allows the generator to incorporate the discriminator's learned results. We refer to this novel architecture as Reinforced-GAN.\footnote{To some extent, this reinforced setting makes our Reinforced-GAN algorithm closely resemble a deep learning-powered EM algorithm.}.              Figure~\ref{fig:GAN} illustrates the structure of both the original GAN and our Reinforced-GAN. By integrating this reinforced link, we embed individual optimization tasks within the generator and equilibrium asset price learning tasks within the discriminator. 

\begin{algorithm}[H]
\caption{Reinforced-GAN Algorithm~\label{algo:Main}}
\begin{algorithmic}
\STATE{\textbf{Input: } fix time discretization {$0=t_0<t_1<\cdots<t_K=T$ with $t_k = kT/K$};}
\STATE{\hspace{38pt} initial position and wealth: $\varphi_{n,t_0}^{\theta^{\texttt{gen}}}=\varphi_{n,0}, W_{n,t_0}^{\theta^{\texttt{gen}}} = W_{n,0}, n\in\mfN$;}
\STATE{\hspace{38pt} terminal value of stock price $S^{\texttt{dis}}_{t_K} =S_T= \mathfrak{S}$;}
\STATE{\hspace{38pt} initialization of parameters $\{ S^{\theta^\texttt{dis}}_{t_0}, \theta^{\texttt{gen}}, \theta^{\texttt{dis}}\}$;}
\WHILE{round $\leq$ \texttt{Round}}
\STATE{\# Train Generator:}
\WHILE{epoch $\leq$ \texttt{Epoch}}
\STATE{sample $\Delta B$ with size as $\texttt{batch\_size}\times (K+1)\times d$  iid 
$\sim\cN(0, \XS{T/K})$
;}
\STATE{call Subroutine~\ref{algo:generator} with $(\mu,\sigma) = F^{\theta^{\texttt{dis}}}$ and current $\theta^{\texttt{gen}}$;}
\STATE{output $\texttt{Loss}_{\texttt{gen}}(\theta^{\texttt{gen}})$ from Subroutine~\ref{algo:generator};}  
\STATE{calculate the gradient of $\texttt{Loss}_{\texttt{gen}}(\theta^{\texttt{gen}})$ with respect to $\theta^{\texttt{gen}}$;}
\STATE{back propagate updates for $\theta^{\texttt{gen}}$ via \texttt{Adam};}
\STATE{epoch ++;}
\ENDWHILE
\STATE{\# Train Discriminator:}
\WHILE{epoch $\leq$ \texttt{Epoch}}
\STATE{sample $\Delta B$,  $\texttt{batch\_size}\times (K+1)\times d$  iid Gaussian random variables with variance $\Delta t$;}
\STATE{call Subroutine~\ref{algo:discriminator} with $(\dot\varphi_n, c_n, Z_n) = F^{\theta^{\texttt{gen}_n}}$ and current ${\theta^{\texttt{dis}}}$;}
\STATE{output $\texttt{Loss}_{\texttt{{dis}}}(\theta^{\texttt{{dis}}})$ from Subroutine~\ref{algo:discriminator};}
\STATE{calculate the gradient of $\texttt{Loss}_{\texttt{{dis}}}(\theta^{\texttt{{dis}}})$ with respect to $\theta^{\texttt{{dis}}}$;}
\STATE{back propagate updates for $\theta^{\texttt{dis}}$ via \texttt{Adam};}
\STATE{epoch ++;}
\ENDWHILE
\STATE{round ++;}
\ENDWHILE
\end{algorithmic}
\end{algorithm}

Our adversarial training procedure with a \textit{reinforcement link} (Algorithm~\ref{algo:Main}). Section~\ref{ssec:generator} and Sectiopn~\ref{ssec:discriminator} respectively detail the {Subroutines}~\ref{algo:generator} and {Subroutines}~\ref{algo:discriminator}  used to invoke the generator and discriminator and the theoretical foundations supporting our approach.
Line $5$ in Algorithm~\ref{algo:Main} represent the reinforced link in our Reinforced-GAN structure. In Section~\ref{ssec:theory}, we present the theoretical guarantee behind our Reinforced-GAN algorithm~\ref{algo:Main}. 


\subsection{Generator for Individual Optimization Problem}\label{ssec:generator}
There is a large body of literature on dynamic portfolio optimization models with trading costs.
Given a generic asset prices dynamics with excess return $\mu$ and volatility $\sigma$, each agent's individual optimization problem can be characterized by a system of FBSDEs, and with specific assumptions on the trading costs, closed-form asymptotic approximations can be obtained~\cite{almgren.chriss.01, almgren.li.16, bayraktar2021asymptotics, guasoni.weber.18,kallsen.muhlekarbe.17,moreau.al.17, muhle2023dynamic, shreve1994optimal,soner.touzi.13}. 

With modern deep learning techniques, the FBSDE solver proposed by~\cite{han2018solving} bypass the need to identify the correct boundary conditions and overcome the curse of dimensionality. In parallel, pioneered by~\cite{buehler2019deep}, various reinforcement learning algorithms are implemented and perform extremely successful in portfolio optimization problems with trading costs. The key idea is to directly parametrize the optimal trading rate and optimize the time discretized analogue of each agent's objective functional~\eqref{eqn:geneeral target}. However, both of these algorithm have drawbacks in practice. For example, FBSDE solver does not scale well when the trading horizon is long or the cross-sectional effect of the stocks is strong.  Deep Hedging algorithms require a huge number of simulated sample paths and usually suffers from underfitting when the time horizon is long. 

In our previous work~\cite{shi2023deep}, to take both the advantages of the deep learning algorithms and the closed asymptotic approximations, we proposed the ST-hedging algorithm, which highly relies on the solution to the frictionless analogue of~\eqref{eqn:geneeral target}. Following similar spirit, we ask the user to specify a reference position for Agent-$n$, denote as $\bar\varphi_{n,t}$, with dynamics
\begin{align}\label{eq:dyn merton}
d\bar\varphi_{n,t} = \bar\mu_{n,t} dt + \bar\sigma_{n,t}dB_t, \qquad\bar\varphi_{n,0}=\varphi_{n,0}.
\end{align}
Here, $\bar\mu_{n,t}$ and $\bar\sigma_{n,t}$ are chosen to depend solely on the \emph{known} market processes, meaning they do not rely on the optimal strategies $\dot{\varphi}_{n,t}$ learned in the generator, or the equilibrium quantities $(S_0, \mu_t, \sigma_t)$ determined in the discriminator. Additionally, we require that the market clears at all time, i.e.
\begin{align}\label{cond:clearing for baseline}
\sum_{n\in\mfN}\bar\varphi_{n,t} = s. 
\end{align}
Instead of using Agent-$n$'s current position $\varphi_{n,t}$, we use the generalized fast variable, $\varphi_{n,t} - \bar\varphi_{n,t}$, representing the deviation from a reference position, as the input for the neural networks. With a well-chosen reference position, the variance of this generalized fast variable is upper bounded, improving scalability as the time horizon increases.
To adapt our model setup to a general reinforcement learning framework, we utilize the $(m+1)$-dim process
\begin{align}\label{def: X}
    X_{n,t} \eqdef (\varphi_{n,t}^\top-\bar\varphi_{n,t}^\top, W_{n,t}^{\dot\varphi})
\end{align}
to denote the state process of Agent-$n$. 
Similarly, we use the $m+1$-dim row vector
\begin{align}
\label{eq:control_ant}
    a_{n,t} 
\eqdef 
    (\dot\varphi_{n,t}^\top,c_{n,t})
\end{align}
to denote the control process of Agent-$n$.
The dynamics of $X$ are thus
\begin{align}\label{dyn:X}
dX_{n,t} =  \mu_{n,t}(X_{n,t},a_t) dt+\sigma_{n,t}(X_{n,t}) dB_t,
\quad X_{n,0} = (0, W_{n,0}).
\end{align}
where the $\mu_{n,t}$ and $\sigma_{n,t}$ are given explicitly by
\begin{align}
\label{eq:Dynamics_mu_and_sigma}
\begin{aligned}
\mu_{n,t}(x,a) &= 
\begin{bmatrix}
\begin{bmatrix}
I_{m\times m} &0
\end{bmatrix}
a^\top-\bar{\mu}_{n,t}\\
(\mu_t^\top, r) x^\top + b_{n,t} - G(\begin{bmatrix}
I_{m\times m} &0
\end{bmatrix}a^\top;\Lambda_t) -(0,1)a^\top
\end{bmatrix},\\
\sigma_{n,t}(x) &= 
\begin{bmatrix}
-\bar\sigma_{n,t}\\
x(\sigma_t^\top,0)^\top + \xi_{n,t}
\end{bmatrix}.
\end{aligned}
\end{align}
And we can further express the objective functional~\eqref{eqn:geneeral target} by $X$ and $a$ with
\begin{align}
\label{eq:Objective_Expecto}
J_n(a) = \EE\left[\int_0^T \tilde{f}_n(t, X_{n,t}, a_t;\mu_t, \sigma_t, \Lambda_t) dt + \tilde{g}_n(X_{n,T})\right].
\end{align}

To help with the design of the discriminator, we need to include a little redundancy by introducing the adjoint BSDE into the generator.  As introduced in~\cite[Chapter 6]{pham2009continuous}, we consider the Hamiltonian for Agent-$n$ as
\begin{align}\label{eqn:hamiltonian n}
\cH_n(t,x,y,z,a)\eqdef \tilde{f}_n(t,x,a;\mu,\sigma,\Lambda) + y^\top\mu_{n,t}(x,a) +\tr(z^\top \sigma_{n,t}(x)).
\end{align}
By the stochastic maximum principle (see~\cite[Chapter 6.4]{pham2009continuous} for details), the optimal trading rate $\dot\varphi_{n,t}$ is related to the $m+1$-dim backward component $Y_{n,t}$
given by the adjoint BSDE: 
\begin{align}\label{eqn: adjoint bsde n}
dY_{n,t} = -\frac{\partial}{\partial x} \cH_n(t, X_{n,t},Y_{n,t},Z_{n,t},a_{n,t})dt +Z_{n,t} dB_t, \quad
Y_{n,T} = \frac{\partial}{\partial x}\tilde{g}_n(X_{n,T}).
\end{align}

With this variance deduction tools and the FBSDE system~\eqref{dyn:X}-\eqref{eqn: adjoint bsde n} on hand, we formulate the learning tasks in the generator.
Consider the time discretizations $0=t_0<t_1<\cdots<t_K=T$, where $t_k = kT/K$ and $\Delta t = T/K$. Let $\left\{\Delta B_{t_k}\right\}_{k=1}^K$ denote an iid normally distributed random variables with mean zero and variance $\Delta t I_d$. For a single simulation, the discretized version of the objective functional~\eqref{eqn:geneeral target} for agent-n can be written as (with a little abuse of notation)
\begin{align}\label{eq:discretized general target}
J_n (a)
= \sum_{k=0}^K \tilde{f}_n(t, X_{n,t_k},a_{t_k}; \mu_{t_k},\sigma_{t_k},\Lambda_{t_k})\Delta t + \tilde{g}_n(X_{n,t_K}).
\end{align}
At the initial time, we use a constant $y_{n,0}^\theta$ to parameterize the initial value of the backward component $Y_{n}$ in the adjoint BSDE~\eqref{eqn: adjoint bsde n}, to simulate the whole adjoint BSDE system forward. At each time $t_k$, we parametrize the control  $a_{n,t_k}^{\theta_{n,k}^{\texttt{gen}}}$, which includes both the trading strategy and the consumption, the initial value $y_{n,0}$ and the volatility $Z_{n,t_k}^{\theta_{n,k}^{\texttt{gen}}}$ of the backward component of the adjoint BSDE~\eqref{eqn: adjoint bsde n} using a neural network $F^{\theta_{n,k}^{\texttt{gen}}}$ with $\tanh$-like activation function as 
\begin{equation}
\label{eq:NNMOdellingAssumption}
(a_{n,t_k}^{\theta_{n,k}^{\texttt{gen}}}, Z_{n,t_k}^{\theta_{n,k}^{\texttt{gen}}})
=
F^{\theta_{n,k}^{\texttt{gen}}}(X_{t_k}^{\theta_{n}^{\texttt{gen}}},B_{t_k}).
\end{equation}
To ease the heavy notation, we denote all involved parameters in the generator by $\theta^{\texttt{gen}} = \{y_{n,0},\theta^{\texttt{gen}}_{n,k}, k=0,1,\ldots, K\}_{n\in\mfN}$. Moreover, all nonlinear activation functions in $F^{\theta^{\texttt{gen}}}$ is $\tanh$. 
Our modeling choice~\eqref{eq:NNMOdellingAssumption} is supported by our main small-time efficient approximability guarantee, which we present in Theorem~\ref{thrm:AprxBaby}. 

The generator's task is therefore to learn the optimal trading strategies of each agent in parallel, where the input of the generator is the simulated Brownian path $\{B_{t_k}\}_{k=0}^K$ and given the dynamics of equilibrium return and volatility $(\mu,\sigma)$. 
With each agent's objective functional, the loss function of the generator can be therefore written as
\begin{align}\label{loss:generator}
\texttt{Loss}_{\texttt{gen}}(\theta^{\texttt{gen}})
\eqdef \sum_{n\in\mfN} 
\left[\left\|Y_{n,t_K}^{\theta^{\texttt{gen}}_n} - \frac{\partial}{\partial x} \tilde{g}(X_{n,t_K}^{\theta^{\texttt{gen}}_n})\right\|^2-J_n(\dot\varphi^{\theta^{\texttt{gen}}_n}_n)\right],
\end{align}
where the first term penalizes the mismatching of the terminal value of the backward component $Y_n$ of Agent-$n$, and the second term is the objective functional. 
We summarize the update procedure of the \textit{generator} in Algorithm~\ref{algo:generator}:

\begin{algorithm}[htp]
\caption{Subroutine: Update Dynamics of Generator~\label{algo:generator}}
\begin{algorithmic}
\STATE{\textbf{Need: }\hspace{2pt} $\Lambda_t, b_{n,t}, \xi_{n,t}, \bar\mu_{n,t}, \bar\sigma_{n,t}$ can be simulated for all $n\in\mfN$;}
\STATE{\textbf{Input: } update rule for $(\mu_{t_k}, \sigma_{t_k})=F^{k}(X_{1,t_k},\ldots, X_{N,t_k}, B_{t_k})$;}
\STATE{\hspace{38pt} parametrization: $(a_{n,t_k}^{\theta_{n,k}^{\texttt{gen}}}, Z_{n,t_k}^{\theta_{n,k}^{\texttt{gen}}})=F^{\theta_{n,k}^{\texttt{gen}}}(X_{t_k}^{\theta_{n}^{\texttt{gen}}},B_{t_k}), n\in\mfN$;}
\STATE{\hspace{38pt} initial value for adjoint backward component $Y_{n,t_0}^{\texttt{gen}} = y_{n,0}^\theta$;}
\STATE{\hspace{38pt} sample path $\Delta B$ with size $\texttt{batch\_size}\times (K+1)\times d$ ;}
\STATE{$X_{n,t_0} = (0, W_{n,0}), J_n= 0$ for each $n\in\mfN$;}
\STATE{$B_{t_0} = 0, k=0$;}
\WHILE{$k\leq K$}
\STATE{for each $n\in\mfN$ in parallel: }
\STATE{\hspace{8pt} update $\Lambda_{t_k}, b_{n,t_k}, \xi_{n,t_k}, \bar\mu_{n,t_k}, \bar\sigma_{n,t_k}$;}
\STATE{\hspace{8pt} $(a_{n,t_k}^{\theta_{n,k}^{\texttt{gen}}}, Z_{n,t_k}^{\theta_{n,k}^{\texttt{gen}}})=F^{\theta_{n,k}^{\texttt{gen}}}(X_{n,t_k}^{\theta_{n}^{\texttt{gen}}},B_{t_k})$;}
\STATE{\hspace{8pt} $J_n\ +=\tilde{f}_n(t_k, X_{n,t_k}^{\theta_{n}^{\texttt{gen}}},a_{n,t_k}^{\theta_{n,k}^{\texttt{gen}}};\mu_{t_k},\sigma_{t_k}, \Lambda_{t_k})\Delta t$;}
\STATE{\hspace{8pt} $\Delta X_{n, t_k}^{\theta_{n}^{\texttt{gen}}}= \mu_{n,t_k} \left(X_{n,t_k}^{\theta_{n}^{\texttt{gen}}},a_{n,t_k}^{\theta_{n,k}^{\texttt{gen}}}\right)\Delta t + \sigma_{n,t_k}(X_{n,t_k}^{\theta_{n}^{\texttt{gen}}})\Delta B_{t_k}$;}
\STATE{\hspace{8pt} $X_{n,t_{k+1}}^{\theta_{n}^{\texttt{gen}}} = X_{n, t_k}^{\theta_{n}^{\texttt{gen}}}+\Delta X_{n, t_k}^{\theta_{n}^{\texttt{gen}}}$;}
\STATE{\hspace{8pt} $\Delta Y_{n,t_{k}}^{\theta^{\texttt{gen}}_n} = - \frac{\partial}{\partial x} \cH_n\left(t_k,X_{n,t_k}^{\theta_{n}^{\texttt{gen}}},Y_{n,t_k}^{\theta_{n}^{\texttt{gen}}},Z_{n,t_k}^{\theta_{n,k}^{\texttt{gen}}},a_{n,t_k}^{\theta_{n,k}^{\texttt{gen}}}\right)\Delta t + Z_{n,t_k}^{\theta_{n,k}^{\texttt{gen}}}\Delta B_{t_k}$;}
\STATE{\hspace{8pt} $Y_{n,t_{k+1}}^{\theta^{\texttt{gen}}_n} = Y_{n,t_k}^{\theta^{\texttt{gen}}_n}+\Delta Y_{n,t_k}^{\theta^{\texttt{gen}}_n}$;}
\STATE{$B_{t_{k+1}} = B_{t_k} + \Delta B_{t_k}$;}
\STATE{$k++$;}
\ENDWHILE
\STATE{for each $n\in\mfN$ in parallel: }
\STATE{\hspace{8pt} $J_n\; += \tilde{g}_n(X_{n,t_K}^{\theta^{\texttt{gen}}_n})$;}
\STATE{\hspace{8pt} $\Delta Y_{n}^{\theta^{\texttt{gen}}_n}=Y_{n,t_K}^{\theta^{\texttt{gen}}_n} - \frac{\partial}{\partial x}\tilde{g}_n(X_{n,t_K}^{\theta^{\texttt{gen}}_n})$;}
\STATE{$\texttt{Loss}_{\texttt{gen}}(\theta^{\texttt{gen}}) = \sum_{n\in\mfN}\left[\|\Delta Y_{n}^{\theta^{\texttt{gen}}_n}\|^2- J_n\right]/\texttt{batch\_size}$;}
\STATE{\textbf{Output: } $\texttt{Loss}_{\texttt{gen}}(\theta^{\texttt{gen}})$ with gradient information.}
\end{algorithmic}
\end{algorithm}

\begin{remark}
This architecture of the generator is designed for the most general case. For specific problems, such as the examples we show in Section~\ref{sec:exp}, we can adjust the structure or the choice of variables for better performance. 
\end{remark}

\subsection{Discriminator for Equilibrium Asset Price Dynamics}\label{ssec:discriminator}
To determine the equilibrium asset price dynamics~\eqref{eq:Sdyn}, 
there are two constraints need to be satisfied: the market clearing condition and the terminal liquidation condition. 
With the optimal control $\{a^{\theta^{\texttt{gen}}_n}_n\}_{n\in\mfN}$ learned from the generator, the learning task in the discriminator is to provide a neural network approximations of the initial stock price $S_0$, the excess equilibrium return $\mu$, and the equilibrium volatility $\sigma$. 

\paragraph{Explicit representation of equilibrium return $\mu_t$} 
In special cases, one is given or can derive the closed-from dependence of the equilibrium return $\mu_t$ on equilibrium volatility $\sigma_t$ and/or the state variables $X_{n,t}$ of each agent, which we express as
\begin{align}\label{eq: mu given}
\mu_t = \mu(t, \sigma_t, \{X_{n,t}\}_{n\in\mfN} ).
\end{align}
Examples of closed-form dependencies include quadratic trading cost models with LQ preferences~\cite{muhle2023equilibrium}, two-agent frictional models with offsetting positions~\cite{shi2020equilibrium},
 or Nash equilibria~\cite{casgrain2022deep,micheli2023closed}.

Although explicit representations are not the main focus of our paper, the \emph{separation} of  for learning optimal trading strategies and equilibrium price dynamics outperforms FBSDE Solvers~\cite{han2018solving, muhle2023equilibrium}. Unsurprisingly, our Reinforced-GAN performs even better with the added dependence in~\eqref{eq: mu given}. See Section~\ref{sec:exp} for details.

\paragraph{Learning of equilibrium return $\mu_t$ with implicit relationship}
Without the closed-from dependence relationship, the equilibrium return $\mu$ is determined implicitly via the market clearing condition in Definition~\eqref{frictional equilibrium} ii), which we use a neural network approximation to parameterize $\mu$.  
The key challenge is designing the discriminator's loss function. 
A natural approach is to penalize the $L_2$ loss of the market clearing condition and the terminal stock price condition. However, this approach fails to converge due to insufficient gradient information for the market clearing condition, limiting the discriminator's performance.

Our approach is inspired by the adjoint BSDE~\eqref{eqn: adjoint bsde n}. Given the convexity of trading costs $G$ and the concavity of $f$ in the objective functional (and consequently $\tilde{f}$ after the variable change), the optimal control for Agent-$n$ can be explicitly expressed via the adjoint variable $Y$ as
\begin{align}\label{eqn: optimal trading strategy by t,X,Y}
a_{n,t} = I_n(t, X_{n,t}, Y_{n,t}; \mu_t, \sigma_t, \Lambda_t)^\top.
\end{align}
where $I_n$ is determined only by $G$ and $\tilde{f}$. 
Here, recall that $a_{n,t}$ is a $(m+1)$-dim row vector. 
With~\eqref{eqn: optimal trading strategy by t,X,Y}, the market clearing condition in Definition~\eqref{frictional equilibrium} ii) can be expressed as, for all $t\in\mfT$, 
\begin{align*}
\sum_{n\in\mfN} 
\begin{bmatrix}
I_{m\times m} &0
\end{bmatrix}
I_{n}(t,X_{n,t}, Y_{n,t}; \mu_t, \sigma_t, \Lambda_t)
&=\sum_{n\in\mfN}
\begin{bmatrix}
I_{m\times m} &0
\end{bmatrix}a_{n,t}^\top
=\sum_{n\in\mfN} \dot\varphi_{n,t}
=0.
\end{align*}
Moreover, to calculate the adjoint backward component $Y_{n}$ more accurately, we use the following expression:
\begin{align}\label{eqn: adjoint Y}
Y_{n,t} = \EE\left[\frac{\partial}{\partial x}\tilde{g}_n(X_{n,T}) + \int_t^T \frac{\partial}{\partial x}\cH_n(u,X_{n,u}, Y_{n,u},Z_{n,u},a_{n,u})du|\cF_t\right].
\end{align}
If $Z_{n,t}$ is not included in~\eqref{eqn: adjoint Y}, the loss function $\texttt{Loss}_{\texttt{gen}}$ of the generator and approximation of $Y_{n}$ can be further simplified. For the same time discretization, at each time $t_k$, we parametrize the equilibrium return and volatility $(\mu^{\theta_k^{\texttt{dis}}}_{t_k}, \sigma^{\theta_k^{\texttt{dis}}}_{t_k})$ using a neural network $F^{\theta_k^{\texttt{dis}}}$, with inputs consisting of the fast variables and simulated Brownian motion $(X_{1,t_k}^{\theta^\texttt{gen}_{1,k}},\ldots, X_{N,t_k}^{\theta^\texttt{gen}_{N,k}}, B_{t_k})$.
In summary, let the parameters of the discriminator be $\theta^{\texttt{dis}} = \{S_{0}^\theta, \theta^{\texttt{dis}}_{k}, k=0,1,\ldots, K\}_{n\in\mfN}$. The update procedure for the discriminator is shown in Algorithm~\ref{algo:discriminator}.

\begin{algorithm}[htp]
\caption{Subroutine: Update Dynamics of Discriminator~\label{algo:discriminator}}
\begin{algorithmic}
\STATE{\textbf{Need: }\hspace{2pt} $\Lambda_t, b_{n,t}, \xi_{n,t}, \bar\mu_{n,t}, \bar\sigma_{n,t}$ can be simulated for all $n\in\mfN$;}
\STATE{\textbf{Input: } update rule $(a_{n,t_k}, Z_{n,t_k})=F^{n,k}(X_{t_k},B_{t_k})$, for  each $n\in\mfN$;}
\STATE{\hspace{38pt} parametrization: $(\mu^{\theta_k^{\texttt{dis}}}_{t_k}, \sigma^{\theta_k^{\texttt{dis}}}_{t_k})=F^{^{\theta_k^{\texttt{dis}}}}(X_{1,t_k},\ldots, X_{N,t_k}, B_{t_k})$;}
\STATE{\hspace{38pt} initial value for stock price $S_{t_0}^{\theta^{\texttt{dis}}} = S_{0}^\theta$;}
\STATE{\hspace{38pt} sample path $\Delta B$ with size $\texttt{batch\_size}\times (K+1)\times d$ ;}
\STATE{$B_{t_0} = 0, J_n(\dot\varphi_n) = 0, k=0$;}
\STATE{\# Forward pass for forward state variable $X_n, n\in\mfN$: }
\STATE{$X_{n,t_0} = (0, W_{n,0})$ for each $n\in\mfN$;}
\WHILE{$k\leq K$}
\STATE{\textbf{if} \emph{expression of $\mu_t$ is known} \textbf{then}}
\STATE{\hspace{8pt} $(\_\_\_\_,\sigma^{\theta_k^{\texttt{dis}}}_{t_k})=F^{^{\theta_k^{\texttt{dis}}}}(X_{1,t_k},\ldots, X_{N,t_k}, B_{t_k})$;}
\STATE{\hspace{8pt} update $\mu^{\theta_k^{\texttt{dis}}}_{t_k}$ via~\eqref{eq: mu given} and group $(\mu^{\theta_k^{\texttt{dis}}}_{t_k}, \sigma^{\theta_k^{\texttt{dis}}}_{t_k})$;}
\STATE{\textbf{else}}
\STATE{\hspace{8pt} $(\mu^{\theta_k^{\texttt{dis}}}_{t_k}, \sigma^{\theta_k^{\texttt{dis}}}_{t_k})=F^{^{\theta_k^{\texttt{dis}}}}(X_{1,t_k},\ldots, X_{N,t_k}, B_{t_k})$;}
\STATE{\textbf{end}}
\STATE{$S_{t_{k+1}}^{\theta^{\texttt{dis}}} = S_{t_k}^{\theta^{\texttt{dis}}} + \mu^{\theta_k^{\texttt{dis}}}_{t_k} \Delta t+\sigma^{\theta_k^{\texttt{dis}}}_{t_k}\Delta B_{t_k}$;}
\STATE{for each $n\in\mfN$ in parallel: }
\STATE{\hspace{8pt} update $\Lambda_{t_k}, b_{n,t_k}, \xi_{n,t_k}, \bar\mu_{n,t_k}, \bar\sigma_{n,t_k}$;}
\STATE{\hspace{8pt} $(a_{n,t_k}, Z_{n,t_k})=F^{n,k}(X_{n,t_k},B_{t_k})$;}
\STATE{\hspace{8pt} $\Delta X_{n, t_k}= \mu_{n,t_k} \left(X_{n,t_k},a_{n,t_k}\right)\Delta t + \sigma_{n,t_k}(X_{n,t_k})\Delta B_{t_k}$;}
\STATE{\hspace{8pt} $X_{n,t_{k+1}} = X_{n, t_k}+\Delta X_{n, t_k}$;}
\STATE{$B_{t_{k+1}} = B_{t_k} + \Delta B_{t_k}$;}
\STATE{$k++$;}
\ENDWHILE
\STATE{\# Backward pass for adjoint backward adjoint component $Y_n, n\in\mfN$:}
\STATE{$k=K$;}
\STATE{$Y_{n,t_K} = \frac{\partial}{\partial x} \tilde{g}_n(X_{n,t_K})$ for each $n\in\mfN$;}
\WHILE{$k\geq0$}
\STATE{for each $n\in\mfN$ in parallel: }
\STATE{\hspace{8pt} $I_{n,t_k} = I_n(t,X_{n,t_k},Y_{t,k};\mu^{\theta_k^{\texttt{dis}}}_{t_k}, \sigma^{\theta_k^{\texttt{dis}}}_{t_k}, \Lambda_{t_k})$;}
\STATE{\hspace{8pt} $Y_{n,t_{k-1}} = Y_{n,t_k}+\ \frac{\partial}{\partial x}\cH_n(t,X_{n,t_k}, Y_{n,t_k},Z_{n,t_k},a_{n,t_k})\Delta t$;}
\STATE{$k--$;}
\ENDWHILE
\STATE{$\texttt{Loss}_{\texttt{dis}}(\theta^{\texttt{dis}}) = \left[\|S_{t_K}^{\theta^{\texttt{dis}}} -\mfS \|^2+\sum_{k=0}^K \|\sum_{n\in\mfN}I_{n,t_k}\|^2\right]/\texttt{batch\_size}$}
\STATE{\textbf{Output: } $\texttt{Loss}_{\texttt{dis}}(\theta^{\texttt{dis}})$ with gradient information.}
\end{algorithmic}
\end{algorithm}
\begin{remark}
Again, this architecture of the discriminaor is designed for the most general case. For specific problems, such as the examples we show in Section~\ref{sec:exp}, we can also adjust the structure or the choice of variables for better performance. 
\end{remark}

\subsection{Theoretical Guarantees}\label{ssec:theory}
Our theoretical guarantees ensure that our approximations between and on discrete-time updates in Algorithm~\eqref{algo:Main} with generator in Subroutines~\ref{algo:generator} and discriminator in Subroutine~\eqref{algo:discriminator} legitimately converges to the true process being approximated. 
To wit, the description of our kernel algorithm for both generator and discriminator are in Appendix~\ref{ssec: kernel}, which consists of a Deep Hedging type algorithm (see~\cite{buehler2019deep}) for a policy iteration to learn the optimal control and an FBSDE solver (see~\cite{han2018solving}). In particular, the convergence analysis for Deep Hedging and FBSDE solver are studied in~\cite{buehler2019deep} and \cite{han2020convergence}, respectively. Hence the convergence of our algorithm is guaranteed. 

Our main theoretical result guarantees that neural network approximations provide a universal and computationally tractable parametric tool for each discretized time increment. Our algorithm is grounded in the following theoretical guarantee: a controlled neural SDE, with control parameterized by a neural network, can approximate the solution to any controlled SDE over a random positive time interval. Furthermore, if this time interval is sufficiently small, the total number of trainable parameters scales linearly, up to polylogarithmic factors, with the reciprocal of the approximation error.

\begin{theorem}[Main Approximation Guarantee]
\label{thrm:AprxBaby}
Fix 
a maximal time discretization step $\Delta T>0$.
Under some regularity condition on the system (i.e. Assumption~\ref{ass:strong solution} - \ref{ass:Polycube} in Appendix~\ref{app:convergence}), we focus on the short period $[t,t+\Delta T]$.
Then for every initialization error satisfying $\EE\left[\|S_t-S^\theta_t\|+\sum_{n\in\mfN}\|Y_{n,t}-y_{n,t}^\theta\|\right]<\varepsilon$ with $0<\varepsilon\le 1$, there exists a constant $c>0$, a stopping time $0<\tau\le \Delta T$ a.s.,  $\tanh$-MLPs $F^{\theta_{\texttt{gen}}}$ and $F^{\theta_{\texttt{dis}}}$, 
\begin{align*}
\mathbb{P}\biggl(\sup_{t\le u \le t+\tau}\left[\|\mu_u-\mu_u^{\theta_{\texttt{dis}}}\| +\|\sigma_u-\sigma_u^{\theta_{\texttt{dis}}}\|+\sum_{n\in\mfN}\|a_{n,u} - a_{n,u}^{\theta_{\texttt{gen}}}\|\right]\leq3\sqrt{\varepsilon}\biggr)\geq 1-c \sqrt{\varepsilon}.
\end{align*}
In particular, $\tau>0$ can be made to be ``small enough'', so that ${F}^{\theta_{\texttt{gen}}}$ and ${F}^{\theta_{\texttt{dis}}}$ need not have more that $\tilde{\mathcal{O}}(1/\varepsilon)$ non-zero trainable parameters.
\end{theorem}
A more general and technical version of Theorem~\ref{thrm:AprxBaby} can be derived, accommodating broader classes of $\tanh$-like activation functions. This version quantitatively captures the effects of any additional smoothness in the dynamics of $\mu$ and $\sigma$ and introduces various technical parameters that can be leveraged. Theorem~\ref{thrm:Aprx}, presented in our paper's Appendix, provides this result, and its proof implies Theorem~\ref{thrm:AprxBaby}.

\section{Ablation Study for LQ Preferences}\label{sec:exp}

This section demonstrates the performance of our algorithm in a frictional market model with randomness driven by a 1-dimensional Brownian motion $B$. We present three examples in Sections~\ref{ss: 2} and~\ref{ss: 3/2}.
We adopt similar settings in~\cite{muhle2023equilibrium,shi2020equilibrium} to ensure comparability. The financial market includes one risky asset with an interest rate of $r=0$. We focus on power trading cost functions:
\[
G(x) = \lambda|x|^q/q,\qquad q\in(1,2].
\]
For the risky asset price dynamics in~\eqref{eq:Sdyn}, the initial stock price $S_0$, equilibrium return $\mu_t$, and volatility $\sigma_t$ are determined, with a terminal liquidation dividend of linear form
\begin{align}\label{terminal: power}
\mfS = \alpha B_T + \beta T, \qquad \mbox{with }\; \alpha,\beta>0. 
\end{align}
The dynamics of  endowment process for Agent-$n$ is assumed to be:
$$
d\zeta_{n,t} = \xi_{n,t} dB_t, \qquad b_{n,t}=0, \qquad \xi_{n,t} = \xi_n B_t,
$$
and with $\bar\gamma\eqdef (\sum_{n\in\mfN}1/\gamma_n)^{-1}$, the initial position of Agent-$n$ is \footnote{Together with the aggregate endowment being zero~\eqref{cond: zero aggregate endowment}, this is also the frictionless initial position of Agent-$n$, see~\cite{herdegen2021equilibrium,muhle2023equilibrium}.}
$
\varphi_{n,0} = \frac{\bar\gamma}{\gamma_n} s$.  
In addition, we assume that the aggregate endowment in this financial market is zero, i.e.
\begin{align}\label{cond: zero aggregate endowment}
\sum_{n\in\mfN} \xi_{n} = 0.
\end{align}
We pick the same LQ preference~\eqref{ex:LQ}, as in e.g.~\cite{garleanu.pedersen.13}, where Agent-$n$ picks the optimal trading rate $\dot\varphi_{n,t} = d\varphi_{n,t}/dt$ to maximize:
\begin{align}\label{target: experiments}
\max_{\dot\varphi\in\cA}J_n(\dot\varphi) 
= \max_{\dot\varphi\in\cA} \EE\left[\int_0^T \varphi_t \mu_t - \frac{\gamma_n}{2} (\varphi_t \sigma_t +\xi_n B_t)^2 -\frac{\lambda}{q}|\dot\varphi_t|^q\; dt\right]. 
\end{align}

\paragraph{Choice of Variables} 
From Proposition 3.3 in~\cite{muhle2023equilibrium} (or Definition~4.1  in~\cite{herdegen2021equilibrium}), with the aggregate endowment being zero,  the frictionless analogue of this financial market has equilibrium volatility and the return 
$$
\bar\sigma_t = \alpha, \qquad \bar\mu_t = \bar\gamma\alpha^2,
$$
hence following equation (3.3) in~\cite{muhle2023equilibrium}, the reference benchmark position for Agent-$n$ can be chosen as the frictionless equilibrium positions:
\begin{align}
\bar\varphi_{n,t} = \frac{\bar\gamma}{\gamma_n} s - \frac{\xi_n}{\alpha}B_t, 
\qquad \bar\varphi_{n,0} = \frac{\bar\gamma}{\gamma_n} s  = \bar\varphi_{n,0}.
\end{align}
Moreover, the wealth $W_{n,t}$ for Agent-$n$ does not show up in the objective functional, so the state variable $X_{n,t}$ for the individual optimization problem simplifies to 
$$
X_{n,t} = \varphi_{n,t} - \bar\varphi_{n,t} = \varphi_{n,t} - \frac{\bar\gamma}{\gamma_n} s + \frac{\xi_n}{\alpha}B_t,
$$
with the dynamics as
$$
dX_{n,t} = \dot\varphi_{n,t} dt + \frac{\xi_n}{\alpha} dB_t.
$$

\paragraph{Simplification of Algorithm~\ref{algo:Main}}
With the above choice of state variables, the Hamiltonian for Agent-$n$ is given  by
\begin{align*}
  \cH_n(t,x,y,z,a) 
& = \left(x+\frac{\bar\gamma}{\gamma_n}s-\frac{\xi_n}{\alpha}B_t\right)\mu_t  - \frac{\gamma_n}{2}\sigma_t^2 \left( x +\frac{\bar\gamma}{\gamma_n}s+\left(\frac{1}{\sigma_t}-\frac{1}{\alpha}\right)\xi_n B_t\right)^2
\\&\qquad    - \frac{\lambda}{q}|a|^q + ay + \frac{z \xi_n}{\alpha}.
\end{align*}
In this case, 
the optimal trading rate $\dot\varphi_{n,t}$ satisfies
$$
0=\frac{\partial}{\partial a} \cH_n(t,X_{n,t},Y_{n,t},Z_{n,t},\dot\varphi_{n,t}) 
= Y_{n,t} - \lambda |\dot\varphi_{n,t}|^{q-1} \sign(\dot\varphi_{n,t}),
$$
which yields the explicit expression as
\begin{align}\label{eq: I for power costs}
\dot\varphi_{n,t} = \sign(Y_{n,t})\left|\frac{Y_{n,t}}{\lambda}\right|^{\frac{1}{q-1}}.
\end{align}
To wit, this relationship reveals that the backward component $Y_{n}$ in the adjoint BSDE is exactly Agent-$n$'s \emph{marginal trading costs} proposed in~\cite{gonon2021asset,shi2020equilibrium}. 
Using the equivalent characterization from the Hamiltonian, the generator adjoint BSDE of Agent-$n$ becomes
\begin{align}\label{ex: BSDE driver}
-\frac{\partial}{\partial x} \cH_n(t,X_{n,t},Y_{n,t},Z_{n,t},\dot\varphi_{n,t}) 
&= -\mu_t +\gamma_n\sigma_t^2 \left(X_{n,t}+\frac{\bar\gamma}{\gamma_n}s+(\frac{1}{\sigma_t}-\frac{1}{\alpha})\xi_n B_t\right)\notag\\
&= \gamma_n \sigma_t (\varphi_t \sigma_t +\xi_n B_t) -\mu_t,
\end{align}
which does not contain the volatility $Z_{n,t}$ of the backward component $Y_{n,t}$.
Together with $\tilde{g}_n(X_{n,T})=0$, we can write the $Y_{n,t}$ process as 
\begin{align}\label{ex: Y for agent n}
Y_{n,t} 
&
= \EE \left[\int_t^T\frac{\partial}{\partial x} \cH_n(t,X_{n,u},Y_{n,u},Z_{n,u},\dot\varphi_{n,u}) du | \cF_t \right]
 \notag\\
 &
=\EE\left[\int_t^T \left( \mu_u-\gamma_n \sigma_u (\varphi_u \sigma_u +\xi_n B_u)\right) du|\cF_t\right]. 
\end{align}
In this case, we may further streamline the update procedure for the generator as well as for the discrimination.  These \textit{light} versions of our algorithms are respectively detailed in Algorithms~\ref{algo:generator for power cost} and~\ref{algo:discriminator for power cost} in our Appendix~\ref{a:AlgosExtras}.

\paragraph{Implementations}
In Section~\ref{ss: 2} with quadratic trading costs, the numerical results obtained by Reinforced-GAN is compared to the closed-from equilibrium solution discussed in~\cite{muhle2023equilibrium}. 
In Section~\ref{ss: 3/2} with superlinear costs of power $3/2$, we first compare our the numerical results by Reinforced-GAN with the leading order approximation of the equilibrium return and volatility from~\cite{shi2020equilibrium} in the 2-agent equilibrium model. Then showcase the potential of our proposed Reinfored-GAN algorithm by the numerical results of a 5-agent equilibrium model, which analytical approach is intractable to the best of our knowledge.

\subsection{Quadratic Trading Costs Equilibrium Models}\label{ss: 2}
When the elasticity parameter $q=2$, it has been well studied that this quadratic trading costs case corresponds to the linear price impact~\cite{garleanu.pedersen.13, garleanu.pedersen.16,herdegen2021equilibrium,muhle2023dynamic}. Notice that with plugging $q=2$ into~\eqref{eq: I for power costs} and combining~\eqref{ex: Y for agent n}, the optimal trading rate $\dot\varphi_{n,t}$ for Agent-$n$ becomes
$$
\dot\varphi_{n,t} = \sign(Y_{n,t}) \frac{|Y_{n,t}|}{\lambda}=\frac{Y_{n,t}}{\lambda} = \EE\left[\int_t^T \left( \mu_u-\gamma_n \sigma_u (\varphi_u \sigma_u +\xi_n B_u)\right) du |\cF_t\right].
$$
Then, the market clearing condition translates to 
\begin{align*}
\lambda \sum_{n\in\mfN} \dot\varphi_{n,t}
= \sum_{n\in\mfN} Y_{n,t}
=\EE\left[\int_t^T \sum_{n\in\mfN}\left( \mu_u-\gamma_n \sigma_u (\varphi_u \sigma_u +\xi_n B_u)\right) du |\cF_t\right],
\end{align*}
which yields the closed-form expression of the equilibrium return:
\begin{align}\label{eq: mu quadratic costs}
\mu_t = \frac{1}{N}\sum_{n\in\mfN} \gamma_n\sigma_t\left(\sigma_t \varphi_{n,t} + \xi_{n}B_t\right). 
\end{align}

Moreover, with arbitrary number of participating agents, \cite{muhle2023equilibrium} has shown that there \emph{exists} a frictional equilibrium solution, given by a system of matrix-valued Riccati ODEs. Hence with quadratic costs, we can compare Reinforced-GAN with the (ground-truth) solution given by the ODE system introduced in~\cite{muhle2023equilibrium}, with and without the update rule for the equilibrium return $\mu_t$ from~\eqref{eq: mu quadratic costs}. 

\begin{figure}[H]
    \centering
       \includegraphics[width=0.45\linewidth]{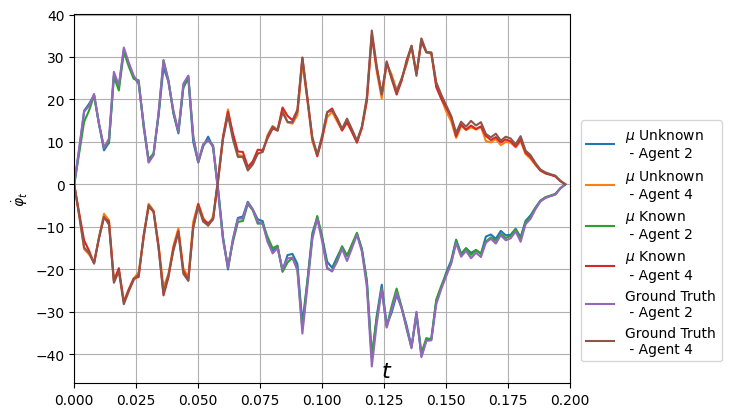}
        \includegraphics[width=0.45\linewidth]{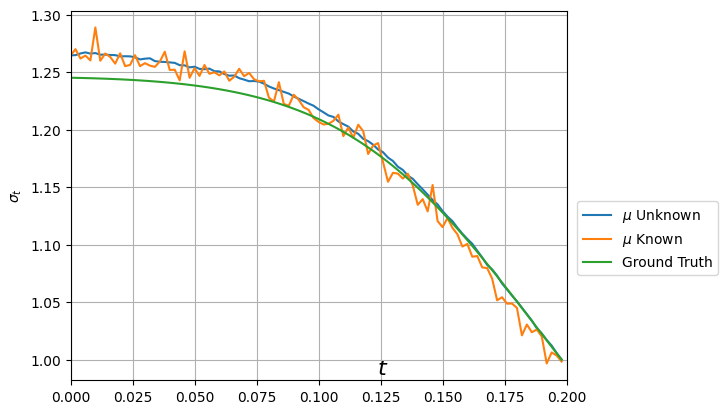}
        \includegraphics[width=0.45\linewidth]{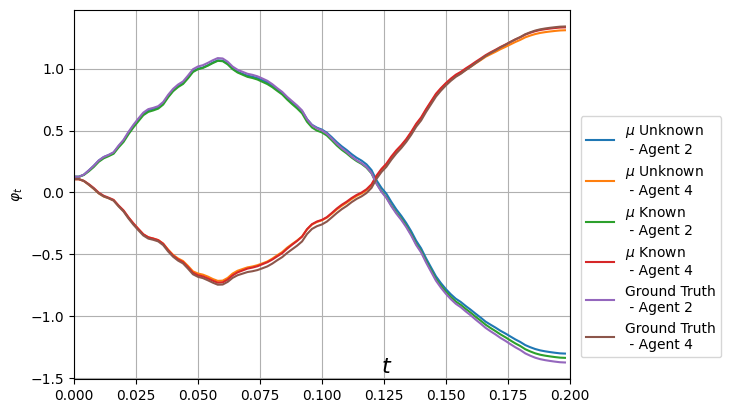}
        \includegraphics[width=0.45\linewidth]{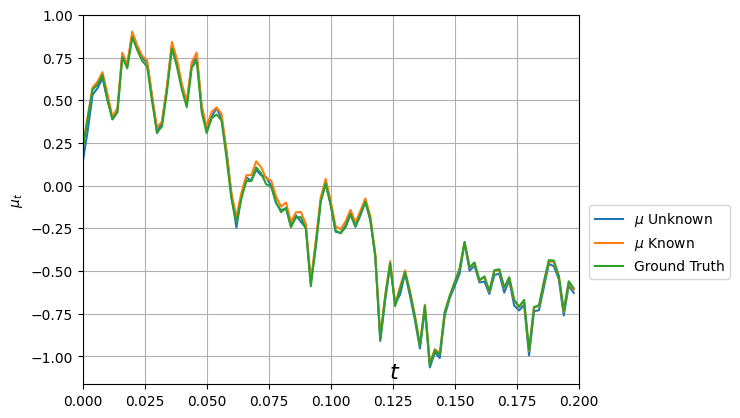}
    \caption{Comparison of Reinforced-GANs Against Ground Truth: 10 Agents with Quadratic Costs. Left panels show a simulation trajectory of Agent-$2$ and Agent-$4$'s optimal trading rates (upper left) and optimal positions (lower left). Right panels show the same simulation trajectory of the equilibrium volatility $\sigma$ (upper right) and equilibrium return $\mu$ (lower right).}
    \label{fig:quad-10}
\end{figure}

\paragraph{10-Agent Frictional Model}
In the first experiments, we consider a frictional market model with agents $N=10$.
The total number of outstanding shares is set as $s=1$, the trading horizon is set as $T=0.2$, the liquidity level parameter is set as $\lambda=0.01$, and the terminal liquidation parameters of~\eqref{terminal: power} are set accordingly as $\beta=2$ and $\alpha=1$. The agents' risk aversion parameters are set as 
$\{1, 1.1, 1.2, 1.3, 1.4, 1.5, 1.6, 1.7, 1.8, 1.9\}$. Moreover, their endowment volatilities are set as $\{28.9, 14.9, 11.8, -14.0, -19.1, -27.0, 22.2, 31.5, -26.3, -22.9\}$, respectively. 
We implement Reinforced-GAN with the generator given by Algorithm~\ref{algo:generator for power cost} and the discriminator by Algorithm~\ref{algo:discriminator for power cost}. In particular, we compare the numerical results of our Reinforced-GANs with and without the dependence of the equilibrium return $\mu_t$ with respect to the equilibrium volatility $\sigma_t$ and the agent's current position $\varphi_{n,t}$ through~\eqref{eq: mu quadratic costs}.
The performance of the generator and discriminator is illustrated in Figure~\ref{fig:quad-10} and summarized in Table~\ref{tab:quad-10}, where we can easily see that our Reinforced-GAN can achieve comparable results to the ground truth,  with or without the dependence relationship~\eqref{eq: mu quadratic costs} of $\mu$.

We start with the performance of the generators, which can be seen from the comparison of the optimal trading rates with respect to the ground truth and the LQ preferences of the agents.
In left panels of Figure~\ref{fig:quad-10}, we plot the optimal trading rates and the corresponding optimal positions of Agent-$2$ and Agent-$4$, where we can see that the numerical results are not far from the ground truth. In particular, at the terminal time $T$, Reinforced-GAN can accurately learn that the optimal trading rate for each agent should be zero  , since it is never optimal to trade if there is no time left for the stock price to change. 
Moreover, the sum of the LQ preferences for all agents learned by Reinforced-GANs is very close to the ground truth, suggesting the success of the generator. Also, it is not surprising to see that with the known dependence relationship~\eqref{eq: mu quadratic costs} of $\mu$, the generator performs slightly better. 

The performance of the discriminators is illustrated by comparing the learned equilibrium return $\mu$ and the learned equilibrium volatility $\sigma$ with the ground truth, the matching of the market clearing condition and the terminal liquidating function, and the initial stock price $S_0$.
In the right panels of Figure~\ref{fig:quad-10}, the learned equilibrium return $\mu$ and the equilibrium volatility $\sigma$ are very close to the ground truth, with or without the dependence relationship of $\mu$ from~\eqref{eq: mu quadratic costs}, indicating that the performance of the discriminator is state-of-the-art. 
For both the market clearing condition and the terminal liquidating condition, Reinforced-GAN achieves almost zero loss, {where these non-zero numbers are largely due to numerical precision}. It is worth noting that \emph{without the dependence relationship~\eqref{eq: mu quadratic costs} of $\mu$}, Reinforced-GAN obtains better numerical results for the initial stock price $S_0$, equilibrium return $\mu$ and the equilibrium volatility $\sigma$, and the market clearing condition and the terminal liquidating condition are better satisfied, illustrating that our design of the discriminator via the adjoint FBSDEs works perfectly for equilibrium models.
\begin{table}[htp!]
	\centering
	\begin{tabular}{|c|c|c|c|c|}
		\hline
		& $\sum_{n\in\mfN} J_n(\dot\varphi_n)$ & $\|\sum_{n\in\mfN}\dot\varphi_n\|^2$ & $\|S_T^\theta - \mfS \|^2$&$S_0$ \\
		\hline
		Ground Truth & $-2.08 \times 10^{-1}$ & $0$ & $0$ & $3.61 \times 10^{-1}$ \\
		\hline
		$\mu$ Known & $-2.09 \times 10^{-1}$ & $2.21 \times 10^{-3}$ & $2.32 \times 10^{-5}$ & $3.58 \times 10^{-1}$ \\
		\hline
		$\mu$ Unknown & $-2.09 \times 10^{-1}$ & $2.30 \times 10^{-5}$ & $2.73 \times 10^{-7}$ & $3.61 \times 10^{-1}$ \\
		\hline
	\end{tabular}
	\caption{Comparison of Reinforced-GANs Against Ground Truth: 10 Agents with Quadratic Costs, simulation is done with 3000 sample paths.}
	\label{tab:quad-10}
\end{table}

\vspace{-20pt}

\subsection{Superlinear Trading Costs Equilibrium Models}\label{ss: 3/2}
To further test our algorithm, we consider the superlinear trading costs with $q=3/2$, which corresponds to the ``square-root'' law as
in~\cite{almgren.chriss.01,lillo.al.03}.

\paragraph{Two-agent frictional market}
When the elasticity parameter $q$ is in $(1,2)$, the optimal trading rate $\dot\varphi_{n,t}$ given by~\eqref{eq: I for power costs} cannot be further simplified. Thus closed-form solution is no longer available. 
However, if there are only two agents in the market, they would take exactly the opposite trading strategy, i.e. $\dot\varphi_{1,t}=-\dot\varphi_{2,t}$. It follows that 
$$
Y_{1,t} = \sign\left(\dot\varphi_{1,t}\right) \left|\dot\varphi_{1,t}\right|^{q-1}
= -\sign\left(\dot\varphi_{2,t}\right) \left|\dot\varphi_{2,t}\right|^{q-1}
=-Y_{2,t}.
$$
Together with~\eqref{ex: Y for agent n} (where the derivation details can be found in~\cite[Chapter 3]{shi2020equilibrium}), we can obtain the closed-form expression for the equilibrium return, i.e.
\begin{align}\label{eq: mu power cost}
\mu_t = \frac{1}{2}\left(\gamma_1\sigma_t\left(\sigma_t \varphi_{1,t} + \xi_{1}B_t\right) + \gamma_2\sigma_t\left(\sigma_t \varphi_{2,t} + \xi_{2}B_t\right)\right).   
\end{align}
Further, there exists a leading order  approximation formula when the trading costs level $\lambda$ is small compared to the trading horizon $T$. 
Details of the derivations can be found in~\cite[Chapter 5]{shi2020equilibrium}.

In this experiment, we keep the total number of outstanding shares as $s=1$, the terminal liquidating parameters as $\beta=2$ and $\alpha=1$, and the level of the trading costs as $\lambda=0.01$. The trading horizon is set as $T=0.4$, in order to apply the leading order approximation. For the agents, we choose their risk aversions as $\gamma_1= 1$ and $\gamma_2=2$, i.e. Agent-$1$ has twice the risk capacity as Agent-$2$, and their endowment volatilities as $\xi_1= 3=-\xi_2$.

\begin{figure}[H]
    \centering
        \includegraphics[width=.45\linewidth]{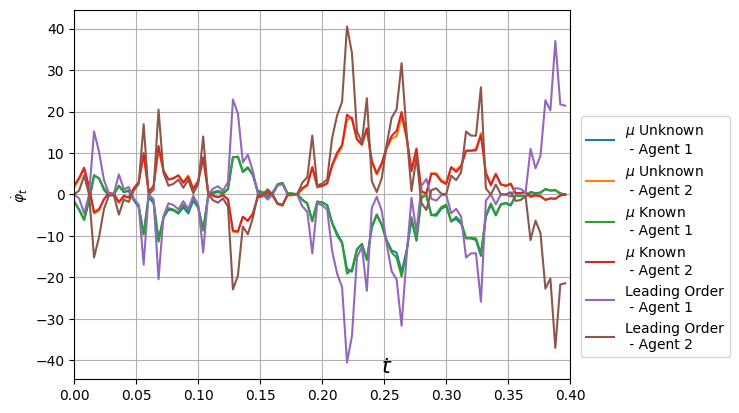}
        \includegraphics[width=.45\linewidth]{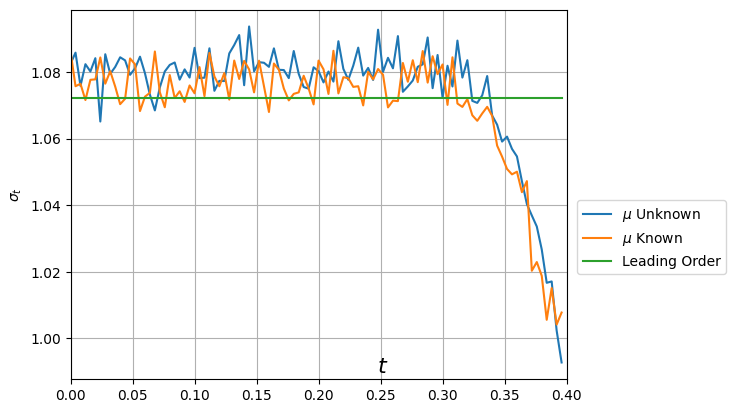}
%
        \includegraphics[width=.45\linewidth]{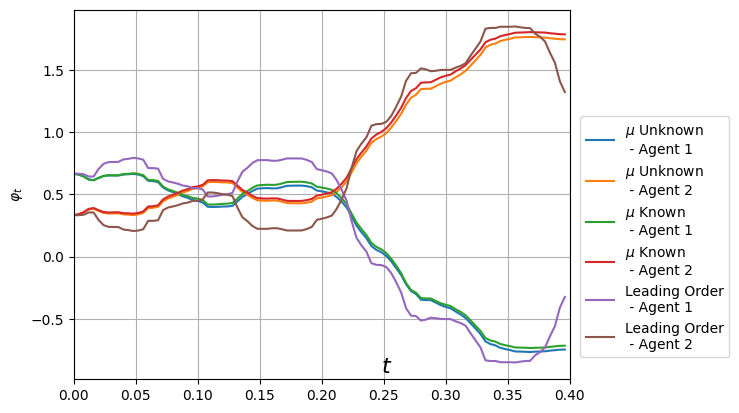}
        \includegraphics[width=.45\linewidth]{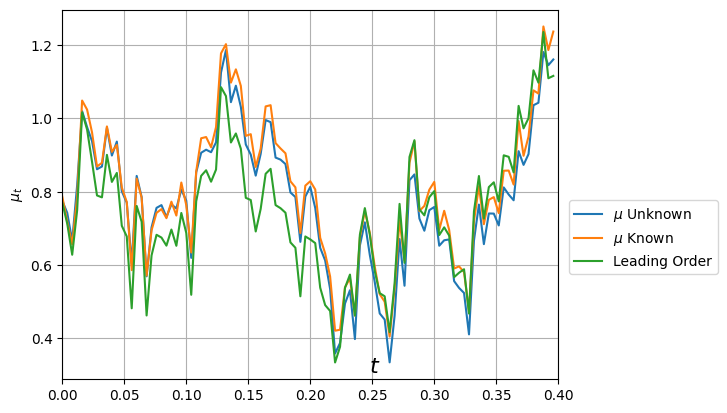}
    \caption{Comparison of Reinforced-GANs Against Leading Order Approximation: Two Agents with $3/2$-Power Costs. Left panels show a simulation trajectory of Agent-$1$ and Agent-$2$'s optimal trading rates (upper left) and optimal positions (lower left). Right panels show the same simulation trajectory of the equilibrium volatility $\sigma$ (upper right) and equilibrium return $\mu$  (lower right).}
    \label{fig:power-2}
\end{figure}

We implemented Reinforced-GAN with and without dependence relationship~\eqref{eq: mu power cost} equilibrium return $\mu_t$ and compared with the leading order approximations provided in~\cite{shi2020equilibrium}. The results are illustrated in Figure~\ref{fig:power-2} and summarized in Table~\ref{tab:power-2}. To start with, the numerical results of the generators provide larger LQ preferences than the leading order approximation. In the upper left panel, the learned optimal trading rates achieve zero at the terminal time, whereas the leading order approximation are still trading actively. 
For the discriminator, we see that both the market clearing condition and the terminal liquidating condition are satisfied. Moreover, in the upper right panel, the learned equilibrium volatility $\sigma$ shows a ``stair-case'' shape, which coincides with the stylized facts. When it is far from the terminal time, the learned volatility $\sigma$ also matches with the leading order approximation, cross-validating the accurateness of the leading order approximation. 
Similarly as in the quadratic costs case in Section~\ref{ss: 2}, 
the discriminators perform even better without this dependence on $\mu$ from the results in Table~\ref{tab:power-2}, where the market clearing condition matches both closer to zero. These observations show that the numerical results learned by the Reinforced-GAN algorithm is a finer approximation to the frictional equilibrium compared to the leading order approximation. When the time to maturity is large compared to the costs parameter $\lambda$, e.g.~\cite[Theorem A.6]{shi2023deep}, the leading order approximation yields similar results comparing to the numerical solution given by the Reinforced-GAN algorithm. When the time to maturity is relatively small, the leading order approximation is no longer accurate comparatively, justifying the usage of the Reinforced-GAN algorithm in this regime. 
\begin{table}[H]
	\centering
	\begin{tabular}{|c|c|c|c|c|}
		\hline
		& $\sum_{n\in\mfN} J_n(\dot\varphi_n)$ & $\|\sum_{n\in\mfN}\dot\varphi_n\|^2$ & $\|S_T^\theta - \mfS \|^2$&$S_0$ \\
		\hline
		Leading Order & $8.94 \times 10^{-4}$ & $0$ & $7.47 \times 10^{-3}$ & $4.15 \times 10^{-1}$ \\
		\hline
		$\mu$ Known & $3.74 \times 10^{-4}$ & $1.07 \times 10^{-2}$ & $2.71 \times 10^{-5}$ & $4.46 \times 10^{-1}$ \\
		\hline
		$\mu$ Unknown & $5.62 \times 10^{-4}$ & $2.19 \times 10^{-6}$ & $1.30 \times 10^{-5}$ & $4.58 \times 10^{-1}$ \\
		\hline
	\end{tabular}
	\caption{Comparison of Reinforced-GANs Against Ground Truth: 2 Agents with 3/2-Power Costs, simulation is done with 3000 sample paths.}
	\label{tab:power-2}
\end{table}

\vspace{-15pt}

\paragraph{More than two agents}

\begin{figure}[H]
    \centering
        \includegraphics[width=.45\linewidth]{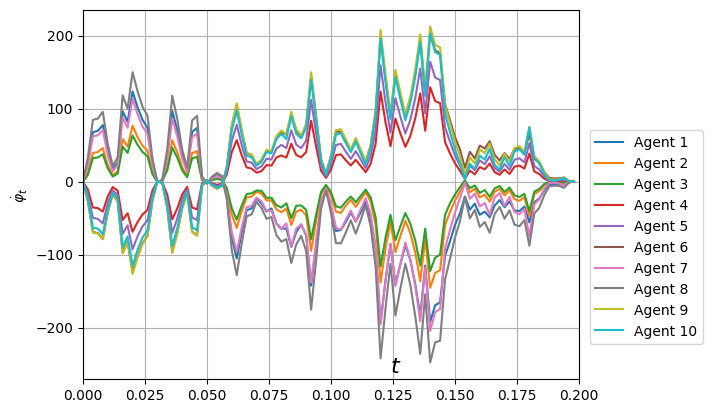}
        \includegraphics[width=.45\linewidth]{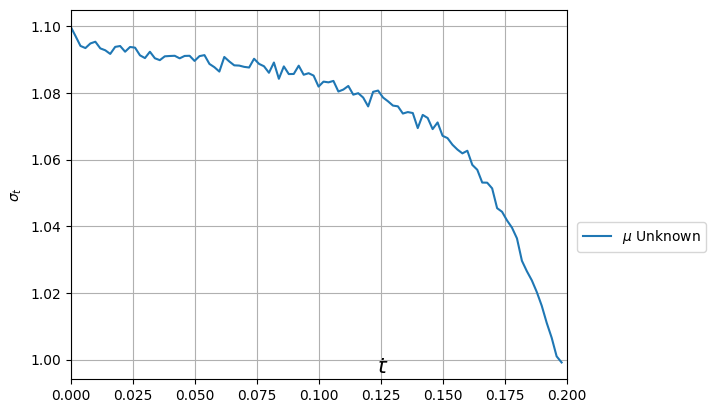}
        \includegraphics[width=.45\linewidth]{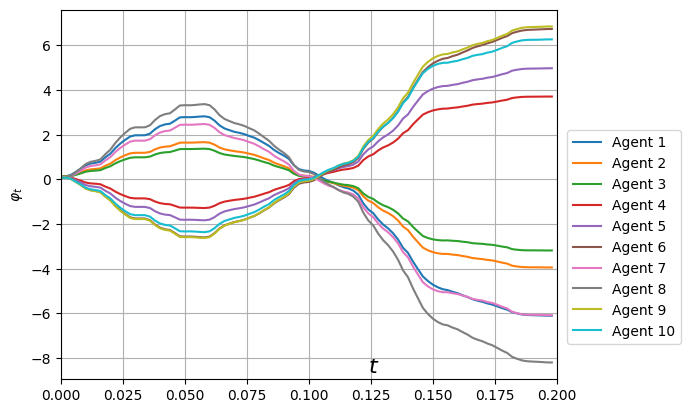}
        \includegraphics[width=.45\linewidth]{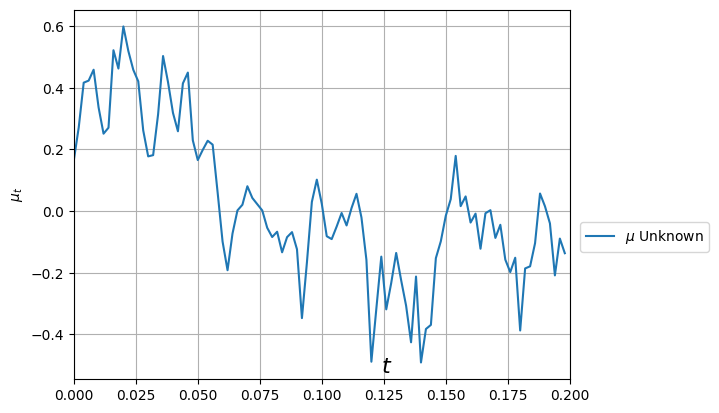}

    \caption{Reinforced-GANs: 10 Agents with 3/2-Power Costs. Left panels show a simulation trajectory of Agent-$1$ - Agent-$10$'s optimal trading rates (upper left) and optimal positions (lower left). Right panels show the same simulation trajectory of the equilibrium volatility $\sigma$ (upper right) and equilibrium return $\mu$  (lower right).}
    \label{fig:power-10}
\end{figure}

{With general power costs and more than two agents, trading among agents becomes complex, making the equilibrium return implicit and preventing leading order approximations. Despite this, the Reinforced-GAN Algorithm~\ref{algo:Main} delivers reliable numerical results that align with stylized facts.
	
This experiment considers a market with 10 agents and $3/2$-power trading costs. For comparison, we set the total number of outstanding shares to $s=1$, the trading horizon to $T=0.2$, and the terminal liquidation parameters as}
$\beta=2$ and $\alpha=1$, and the level of the trading costs as $\lambda=0.01$. For the agents, we choose their risk aversions and their endowment volatility as the same in the quadratic costs case. The implementation results are shown in Figure~\ref{fig:power-10} and Table~\ref{tab:power-10}. 

As shown in the upper left panel, the generator learns that all agents stop their trading at the terminal time $T$. With the same trading costs level but different elasticity parameters, the agents trade more extensively with $3/2$-costs compared to quadratic costs. The market clearing condition and terminal liquidating condition are satisfied, suggesting that the discriminator learns the equilibrium stock dynamics. 
In the upper left panel, the equilibrium volatility shows a shape similar to $\tanh$ as the in Figure~\ref{fig:power-10}, which matches the stylized facts we have of the equilibrium volatility. 
With the same set of parameters for the risk aversions $\gamma$ and the endowment volatilities $\xi$ of the agents, and the same trading horizon $T$, terminal liquidation parameters $\alpha$ and $\beta$, and the trading costs parameter $\lambda$, we can see that the initial stock price $S_0$ for $3/2$-costs is $0.365$, which is larger than the initial stock price $S_0=0.361$ for the quadratic costs case. Given that $3/2$-costs penalize the trading less than quadratic costs when the deviation from the frictionless position is large, the initial stock price is discounted less compared to the quadratic costs case. 

\begin{table}[H]
	\centering
	\begin{tabular}{|c|c|c|c|c|}
		\hline
		& $\sum_{n\in\mfN} J_n(\dot\varphi_n)$ & $\|\sum_{n\in\mfN}\dot\varphi_n\|^2$ & $\|S_T^\theta - \mfS \|^2$&$S_0$ \\
		\hline
		$\mu$ Unknown & $-9.46 \times 10^{-2}$ & $9.32 \times 10^{-5}$ & $5.04 \times 10^{-6}$ & $3.65 \times 10^{-1}$ \\
		\hline
	\end{tabular}
	\caption{Reinforced-GANs: 10 Agents with 3/2-Power Costs, simulation is done with 3000 sample paths.}
	\label{tab:power-10}
\end{table}

\section{Conclusion}\label{sec:con}

In conclusion, our paper presents a tractable computational framework for computing market equilibrium asset pricing in the presence of trading costs. We extend the tractability of traditional models to more realistic settings beyond the LQ setting, including those with stochastic liquidity and time-varying trading costs. We demonstrate how the excess equilibrium return can be derived in two-agent markets and discuss the challenges of scaling to multi-agent systems due to the complexity of the resulting FBSDEs. Our work paves the way for future research to develop tailored numerical methods and explore general frameworks for financial equilibria with trading frictions.  Our empirical results were guided by theoretical approximability guarantees supporting the fact that a \textit{small} neural network approximation of our market equilibria is possible. 

\appendix
\section{Adjoint BSDEs for Individual Optimizations}\label{app: adjoint FBSDE} 

First, using~\cite[Chapter 6]{pham2009continuous}, we can infer that the optimal control $a$ satisfies
\begin{align}\label{I: optimal control}
a_{n,t} &= \max_a  \cH_n(t, X_{n,t},Y_{n,t},Z_{n,t},a)
\notag\\&= \max_{a} \left\{ \tilde{f}_n(t,X_{n,t},a;\mu_t,\sigma_t, \Lambda_t) + Y_{n,t}^\top\ \mu_{n,t}(X_{n,t},a) +\tr(Z_{n,t}^\top \sigma_{n,t}(X_{n,t}))\right\}.
\end{align}
The concavity of $\tilde{f}$ in $a$ is inherited from the concavity of $f$ in $(\dot\varphi^\top,c)$, and by the definition of $\mu_{n,t}$ and $\sigma_{n,t}$ from~\eqref{eq:Dynamics_mu_and_sigma} and the convexity of the cost functional $G$ in $\dot\varphi$, we can see that $\mu_{n,t}$ is also concave in $a$. Therefore, to obtain the existence of an explicit expression of $a$, we need to analyze the property of $Y_{n,t}$ from the adjoint FBSDE. To this end, it is easier for us to write the backward component (with generic input) as follows:
\begin{align*}
Y_{n,t} =
\begin{bmatrix}
Y_{n,t}^{\varphi}\\
Y_{n,t}^W
\end{bmatrix},
\end{align*}
and they satisfies
\begin{align*}
dY_{n,t}^\varphi 
&= -\left(\frac{\partial}{\partial \varphi}f_{n,t} + Y_{n,t}^{W}\mu_t + \tr(\sigma_t^\top Z_{n,t}^{W})\right)dt + Z_{n,t}^\varphi dB_t, && Y_{n,T}^\varphi = \frac{\partial }{\partial x} g_{n,T}, \\
dY_{n,t}^{W}
&=-\left(\frac{\partial}{\partial w}f_{n,t} + r Y_{n,t}^{W} \right)dt + Z_{n,t}^W dB_t, && Y_{n,T}^W= \frac{\partial }{\partial w} g_{n,T}.  
\end{align*}
where we write  $f_{n,t} = f_n(t, W^{\dot\varphi,c}_{n,t}, {\varphi}_{n,t}, \dot\varphi_t,c_t; \mu_t, \sigma_t, \Lambda_t)$  and $g_{n,T} = g_n(\varphi_{n,T},W_{n,T})$ for simplicity.  
Notice that, the argument $\dot\varphi_t$ and $c_t$ are not referring to the optimal trading rate and consumption, but for two generic integrable process. 
We can then turn to the analysis of the optimal consumption and optimal trading strategies.

First, we write the following implicit functions for the optimal consumption $c_{n,t}$:
\begin{align}
c_{n,t} = \max_{c} \left[f_n(t, W^{\dot\varphi,c}_{n,t}, {\varphi}_{n,t}, \dot\varphi_t,c; \mu_t, \sigma_t, \Lambda_t)
- Y_{n,t}^W \ c \right].
\end{align}
Since $f_n$ is strictly concave in c, hence adding a linear function of c will still makes the target function still strictly concave and the explicit expression of $c$ is guaranteed, and it only depends on $X_{n,t}$ and $Y_{n,t}^W$, not the volatility of the backward component.

For the optimal trading rates $\dot\varphi_{n,t}$, 
$$
\dot\varphi_{n,t} = \max_{\varphi_t}\left[
f_n(t, W^{\dot\varphi,c}_{n,t}, {\varphi}_{n,t}, \dot\varphi_t,c; \mu_t, \sigma_t, \Lambda_t)
- Y_{n,t}^W  G(\dot\varphi; \Lambda_t) + Y_{n, t}^\varphi \dot\varphi \right]. 
$$
Given the strict concavity of both $f_n$ and $-G$ in $\dot\varphi$, the explicit expression of $\dot\varphi$ is guaranteed if $Y_{n,t}^W$ remains nonnegative on $[0,T]$, which is provided by Proposition~\ref{prop: positive Y^W_n}. 
Next we present the proof of Proposition~\ref{prop: positive Y^W_n}
to complete the analysis. 
\begin{proposition}\label{prop: positive Y^W_n}
Consider the following FBSDE~\eqref{eq: BSDE for Y^W_n}
\begin{align}\label{eq: BSDE for Y^W_n}
dY_t = -\left(f^Y_t +r Y_t\right)dt+ Z_t dB_t, \qquad Y_T =  g^Y_T. 
\end{align}
With given (generic) integrable process $\varphi_t$, $W_t$, $\dot\varphi_t$ and $c_t$, and the processes $f_t^Y$ and $g^Y_T$ satisfy
$$
f^Y_t = \frac{\partial}{\partial w} f_n(t, \varphi_t, W_t, \dot\varphi_t, c_t; \mu_t, \sigma_t, \Lambda_t)\geq0, \qquad
g^Y_T =  \frac{\partial }{\partial w} g_n(\varphi_{n,T},W_{n,T}) \geq 0,
$$
where the positivity is guaranteed by the definition of $f_n$ and $g_n$.
Then~\eqref{eq: BSDE for Y^W_n} admit an $L^2$ solution and $Y_t^W$ remains nonnegative for all $t\in[0,T]$. 
\end{proposition}
\begin{proof}

The existence and uniqueness of the BSDE is provided by the martingale representation theorem.
To wit, notice that
$$
d e^{rt}Y_t
= -e^{rt} f^Y_t dt  + e^{rt} Z_t dB_t,
$$
then, by the fact that $e^{rT} g_T^Y$ is also integrable, 
we can rewrite $Y_t$ as 
$$
Y_t = e^{-rt}\EE_t[\int_t^T e^{ru} f^Y_u du + e^{rT} g_T^Y] \geq 0
$$
thus concluding our proof.
\end{proof}

\begin{remark}
In traditional methods, establishing the existence and uniqueness of the FBSDE system is challenging due to the coupling between the equations. Our algorithm addresses this by treating each equation individually, viewing the BSDEs in a decoupled manner. This approach simplifies the process, making it straightforward to obtain existence and uniqueness through the martingale representation theorem.  We do not verify whether the optimal solution learned by the generator is indeed represented by the relationship in~\eqref{eqn: optimal trading strategy by t,X,Y}. Therefore, in the generator, all adjoint BSDEs are treated as decoupled.
\end{remark}
\section{Convergence Analysis for Reinforced-GANs}\label{app:convergence}


\subsection{Kernel Algorithm}\label{ssec: kernel}
We identify the kernel algorithm used in both the generator and the discriminator. Notice that the structures of generator and discriminator share the same spirit. 
To generalize the problem, 
we consider a system with forward component $(X_t)_{t\in\mfT}$, control $(a_t)_{t\in\mfT}$, and backward component $(Y_t)_{t\in\mfT}$, with the following dynamics
\begin{align}
dX_t &=\mu^X(t, X_t, a_t, B_t) dt +\sigma^X(t, X_t, B_t) dB_t, && X_0 = x\in\RR^{d_x}, \label{kernel: FSDE}\\
dY_t &= \mu_t^Y(t, X_t, Y_t, Z_t, a_t, B_t) dt + Z_t dB_t
&& Y_T = \mfY. \label{kernel: BSDE}
\end{align}
The target of the problem is to maximize the following target via choosing the control $(a_t)_{t\in\mfT}$:
\begin{align}\label{target: kernel}
J_{\texttt{con}}(a) : = \EE\left[\int_0^Tf(t, X_t, a_t)\ dt + g(X_T)\right],
\end{align}
and solve the (adjoint) BSDE~\eqref{kernel: BSDE}. In other words, we can formulate the problem in the following steps: first use target~\eqref{target: kernel} to find the optimal control $a_t$, and with this optimal control $a_t$ and the associated controlled forward process $X$, we obtain the backward components $(Y, Z)$ from  the BSDE~\eqref{kernel: BSDE}. 

\begin{remark}
For  the generator described in Section~\ref{ssec:generator}, the $X_t$ contains the exogenous components $\Lambda_t, \{\xi_{n,t}, b_{n,t}, \bar\mu_{n,t}, \bar\sigma_{n,t}\}_{n\in\mfN}$, the input $(\mu_t, \sigma_t)$ from the discriminator, and the forward state variable from each Agent-$n$, i.e., $\{X_{n,t}\}_{n\in\mfN}$; the control $a_t$ contains each Agent-$n$'s action $\{a_{n,t}\}_{n\in\mfN}$; the backward variables $(Y_t, Z_t)$ contains the adjoint backward varibale for each Agent-$n$, i.e., $\{(Y_{n,t}, Z_{n,t})\}_{n\in\mfN}$. The $J_{\texttt{con}}(a)$ in the generator is therefore $\sum_{n\in\mfN}J_n(a)$ with $J_n(a)$ defined in~\eqref{eq:Objective_Expecto}.

For the discriminator described in Section~\ref{ssec:discriminator}, the $X_t$ contains the exogenous components $\Lambda_t, \{\xi_{n,t}, b_{n,t}, \bar\mu_{n,t}, \bar\sigma_{n,t}\}_{n\in\mfN}$ and the input $\{X_{n,t}, Y_{n,t}\}$ from the generator; the control $a_t$ corresponds to the equilibrium return $\mu_t$; and the backward variables $(Y_t, Z_t)$ is the equilibrium stock price and volatility $(S_t,\sigma_t)$. 
The $J_{\texttt{con}}(a)$ in the discriminator is in fact the (transformed) market clearing condition, i.e.
to find $\mu_t$ and maximize $- \EE\left[\int_0^T\left\|\sum_{n\in\mfN}I_n(t, X_{n,t},Y_{n,t};\mu_t,\sigma_t,\Lambda_t)\right\|^2\right]$, where $I_n(t, X_{n,t},Y_{n,t};\mu_t,\sigma_t,\Lambda_t)$ is defined as the optimizer functional in~\eqref{I: optimal control}.
\end{remark}

In the kernel algorithm, we fix a time discretizations $0=t_0<t_1<\cdots<t_K=T$, where $t_k = kT/K$ and $\Delta t = T/K$. At time 0, we parametrize the initial value $y_0$ for the backward component. At each time $t$, a shallow neural network $F_t^\theta$ is used to approximate the action $a_t$ and the backward component's volatility $Z_t$, i.e.
$$
(a_t^\theta,Z_t^\theta) = F^\theta_t(X_t, B_t), 
$$
With the initial value $y_0$, the dynamics of the forward component $X$ and backward component $Y$ can then be simulated forward.
The loss function of the networks is therefore
\begin{align}\label{target: kernel loss}
\texttt{Loss}
= \EE[\|Y_T -\mfY\|^2] - J_{\text{con}}(a)    
\end{align}
With the above settings, it is not hard to see that our kernel algorithm for both the generator and the discriminator in our Reinforced-GANs are combinations of the ST-Hedging from~\cite{shi2023deep} (or equivalently the Deep Hedging algorithm from~\cite{buehler2019deep}) and the Deep BSDE Solver from~\cite{han2018solving}. 
Therefore, the existence of the optimizer and the corresponding convergence of our algorithm is guaranteed under the assumptions of the Deep Hedging algorithm from~\cite{buehler2019deep} and the assumptions of the convergence analysis for Deep BSDE Solver from~\cite{han2020convergence}.

\subsection{Convergence Analysis}\label{ssec: assumptions}
In this section, we set up the platform to establish the understanding of why a shallow network works at each time discretization. We start with the introduction of the sigmoidal activation function, and the requirement of the regularity of the system. Then, we present the theoretical guarantees for the approximation, where Theorem~\ref{thrm:AprxBaby} follows as a corollary. 

\paragraph{Activation Function}
In the convergence analysis, we consider sigmoidal activation functions, which are similar to the original (qualitative) universal approximation theorem of~\cite{hornik1989multilayer} but tend to be more \textit{numerically stable} in numerical experiments. Unlike the name suggests, sigmoidal activation functions is a relatively large family of smooth activation functions, including the $\tanh$ activation functions. In addition, we require a second-order non-degeneracy condition of~\cite{zhang2024deep}; further restricting the first-order non-degeneracy condition considered in~\cite{kidger2020universal,kratsios2022universal}.  Note that the \textit{global} approximation properties of networks built using these activations were recently (qualitatively) considered in~\cite{van2024noncompact}.
\begin{definition}[Non-Degenerate Sigmoidal Activation Function]
\label{def:ShapedNonDegen}
An map $\rho:\mathbb{R}\to\mathbb{R}$ is a \textit{non-degenerate sigmoidal} activation function if: $\rho$ is Lipschitz and $\sup_{u\in \mathbb{R}}|\rho(u)|<\infty$ and
\begin{enumerate}
    \item[(i)] \textbf{Sigmoidal:} $\lim\limits_{u\to-\infty}\rho(u)$ and $\lim\limits_{u\to\infty}\rho(u)$ both exist, are finite, and distinct,
    \item[(ii)] \textbf{Non-Degenerate:} the is a $u_0\in \mathbb{R}$ at which $\rho$ is twice differentiable and $\partial^2\rho(u)\neq 0
    $
    .
\end{enumerate}
\end{definition}

\begin{remark}
By the Mean value theorem, $\tanh$ is Lipschitz, and it satisfies Definition~\ref{def:ShapedNonDegen} (i); and $\partial^i\tanh(1)\neq 0$ for $i=0,1,2$.
\end{remark}

\paragraph{Assumptions}

To facilitate the convergence analysis for small time duration, we consider the process $E$ that groups the forward component $X$ and the backward component $Y$ together, i.e., $E_t=(X_t, Y_t)$. Similarly, we group the parameterized control and backward component's volatility, i.e., $\alpha_t=(a_t, Z_t))$. With a little abuse of notation, we use $\mathcal{A}$ to represent the admissible set for $\alpha$. For the parametrized network, we focus on multi-layer perceptrons (MLPs) with sigmoidal activation function,  such as $\tanh$, since they tend to be more numerically stable than ReLU networks in experiments. Further we require the following assumptions on the process $E$:

\begin{assumption}\label{ass:strong solution}
[Strong solution]
Let $d_E,d_\alpha, s$ be positive constants, and  $\mu^E \in C^s(\mathbb{R}^{+d_E+d_\alpha+d}, \mathbb{R}^{d_E})$, 
and 
$\sigma^E\in C^s(\mathbb{R}^{1+d_E+d_\alpha+d}, \mathbb{R}^{d_E\times d})$ 
be smooth functions with Lipschitz constant $L_{\mu^E}\ge 0$ and $L_{\sigma^E}>0$ respectively. Recall that $B_t$ is a $d$-dim Brownian motion. Therefore, we are considering the process $E$ being the unique strong solution for the following SDE with respect to a generic process $\alpha\in\mathcal{A}$: 
\begin{align}\label{eq:SDE_X__true}
dE_t = \mu^E(t,E_t, \alpha_t, B_t) dt + \sigma^E(t, E_t, \alpha_t, B_t)dB_t, \qquad E_0 = e_0.
\end{align}
\end{assumption}

\begin{assumption}\label{ass:recurrency}
[Recurrency] The solution $E$ for~\eqref{eq:SDE_X__true} is a recurrent diffusion. 
\end{assumption}
\begin{assumption}\label{ass:Polycube}
[Polynomially-Bounded Average Exit-Time from Hypercubes]

There exist constant $q>0,c_+>c_{-}>0,$ and $M_0>0$ such that
for each $\alpha\in \mathcal{A}$ and $0<M<M_0$, 
\begin{align*}
c_- M^q \leq \mathbb{E}[\tau_M^E] \leq c_+ M^q 
\end{align*}
where, $\tau_M^E= \inf\{t>0:
E_t\not\in [-M,M]^d
\}$, where $E$ is the strong solution as in~\eqref{eq:SDE_X__true}. 
\end{assumption}

Finally, with a little abuse of notations for smooth function $f$ and $g$, we rewrite the loss function as 
\begin{align}\label{target: alpha}
\texttt{Loss}(\alpha_t)
=  \EE\left[\int_0^T f(t, E_t, \alpha_t) \ dt + g(E_T)\right],   
\end{align}

The assumption~\ref{ass:strong solution} guarantees that for every
admissible strategy $\alpha\in\mathcal{A}$, the associated process $E$ following dynamic~\eqref{eq:SDE_X__true} is well defined. 

A direct consequence of Assumption~\ref{ass:recurrency} is that, for every $\varepsilon>0$ and every $\alpha\in\mathcal{A}$, there exists a constant $M_{\varepsilon, \alpha}>0$ such that for the process $E$ follows~\eqref{eq:SDE_X__true} with respect to this $\alpha$, $\mathbb{P}[|E_t|>M_{\varepsilon,\alpha}]<\varepsilon$. Therefore, we can focus on a relatively large $M_0>0$ such that with high probability $|E_t|<M_0$. In numerical experiments, a typical training protocol is that, if the controlled process $E$ for a (neural network approximated control) $\alpha$ has reached a very large value that is almost beyond the numerical precision's capacity, then one would stopped this run and start a new one. 

In addition, with Assumption~\ref{ass:strong solution} and Assumption~\ref{ass:recurrency}, we can comfortably rest assured that the controlled problem~\eqref{target: alpha} has an optimizer $\alpha^*\in\mathcal{A}$, with high probability. Moreover, the optimizer $\alpha^*$ is a smooth function of the same smooth order as $\mu^E$ and $\sigma^E$, and $\alpha_t^*=\alpha^*(t,E_t^*,B_t)$. We also denote the controlled process with respect to $\alpha^*$ to be $E^*$, and $e^*_0$ for the corresponding initial value of the process $E^*$. 

Finally, our approximation guarantee requires an additional regularity condition on the mean exit time of the controlled process $E$ following~\eqref{eq:SDE_X__true}, hence Assumption~\ref{ass:Polycube} ensures that $E$ exits any small cube centered at the origin with a tight polynomial rate. 

\paragraph{Setup of Neural Networks}
Let $\hat\alpha$ be an $\rho$-MLP with non-degenerate sigmoidal activation function $\rho$ and initializing condition  $\hat{e}_0$ be a random variable. Let $\hat{F}$ denote the mapping of $(t,e,b) \to (\mu^E(t,e, \hat\alpha(t,e,b),b), \sigma^E(t,e, \hat\alpha(t,e,b),b), \hat\alpha(t,e,b))$. 
Since $\rho$ is Lipschitz, affine maps are Lipschitz, and the composition of Lipschitz functions is again Lipschitz, then there is always a unique strong solution to the SDE
\begin{equation}
\label{eq:SDE_X__approx}
d\hat{E}_t  = {\mu}^E(t,\hat{E}_t,\hat{\alpha}_t,B_t)dt + {\sigma}^E(t,\hat{E}_t,\hat{\alpha}_t,B_t)dB_t,
\end{equation}
and with a slight abuse of notation, the \textit{approximate controlled} $\hat{\alpha}_t$ is 
\begin{equation}
\label{eq:SDE_X__controlledprocess}
        \hat{\alpha}_t 
    = 
        \hat\alpha(t,\hat{E}_t,B_t).
\end{equation}

\paragraph{Convergence Analysis}


To begin with, our approximation guarantee provides small time \textit{approximation rates} for \textit{controlled neural SDEs}; namely, objects of the form~\eqref{eq:SDE_X__approx} to objects of the form~\eqref{eq:SDE_X__true}.  We emphasize that the controlled neural SDEs in our guarantee are \textit{light}, in the sense that they converge at a linear rate, up to negligible polylogarithmic factors, in the reciprocal approximation error.   We make use of small randomized time horizons on which our approximation guarantee holds, which allows us to maintain controlled neural SDEs depending only on a few non-zero (trainable) papers.   
Without loss of generality, we focus on $[0,T]$ with the initial state for $E$ is a random variable. Then argument for the (small network) approximation guarantee on period $[t, t+T]$ with ``initial condition''for $E$ at time $t$ follows similarly. 

\begin{lemma}[{Approximation by Controlled SDEs with Correct Initial Condition}]
\label{lem:Universality_Neural_SDE}
Fix a non-degenerate sigmoidal activation function $\rho$ and suppose Assumption~\ref{ass:strong solution} - \ref{ass:Polycube} hold. Moreover, suppose $\hat{e}_0 = {e}_0^*$.  
For every approximation error $\varepsilon>0$ and failure probability $\delta>0$, there exists a $\rho$-MLPs $\hat\alpha$ and an almost surely positive stopping time $\tau_M$ with $0<\mathbb{E}[\tau_M]\leq O( \min\{T,M^q\})$, such that the processes $\hat{E}_t$ in~\eqref{eq:SDE_X__approx} with $\hat\alpha$, satisfies
\begin{equation}
\label{ieq:approx_SDE}
        \mathbb{P}\biggl(
            \sup_{0\le t \le \tau}
                \|E_t^*-\hat{E}_t\|
                +
                \|\alpha_t^{\star}-\hat{\alpha}_t\|
            \le 
                \varepsilon
        \biggr)
    \ge 
        1- \delta
.
\end{equation}
\textbf{Light Networks on Small Times:}
Moreover, $\tau$ can be chosen to be ``small enough'', i.e.\ $0<\mathbb{E}[\tau]\leq O( \varepsilon^{q(1-{s}/{2(1+d_E + d)})})$, so that $\hat{F}$ need not have more that $\tilde{\mathcal{O}}(1/\varepsilon)$ non-zero (trainable) parameters.
\end{lemma}

\begin{lemma}[Perturbation to Initial Conditions]
\label{lem:Simple_Pertrubation_Lemma}
Let $0<\varepsilon < 1$, and random variable $\tilde{e}_0$, with $\EE[\|e_0-\tilde{e}_0\|]\le \varepsilon$, 
and let $\hat{E}$ be the strong solution to the stochastic differential equation~\eqref{eq:SDE_X__true} with a generic control $\alpha$  and initial condition $\tilde{E}_0 = \tilde{e}_0$.
Then, for every $\zeta>0$ we have the following concentration inequality
\begin{equation}
\label{eq:concentration_perturbation_IC}
    \mathbb{P}\biggl(
        \sup_{0\le t\le T}\,
            \|E_t-\tilde{E}_t\|
        \le
            \zeta
    \biggr)
\ge 
    1
    -
    \frac{c_{T,\mu^E,\sigma^E}\varepsilon}{\zeta}.
\end{equation}

\end{lemma}

It remains to deduce the validity of Theorem~\ref{thrm:Aprx}.  This is a direct combination of our approximation result in Lemma~\ref{lem:Universality_Neural_SDE} and our perturbation result in Lemma~\ref{lem:Simple_Pertrubation_Lemma}.  The following is the general form of the result in the main body of our text, namely Theorem~\ref{thrm:AprxBaby}.

\begin{theorem}[Main Approximation Guarantee (General Version)]
\label{thrm:Aprx}
Fix a non-degenerate sigmoidal activation function $\rho$, and a maximal time-horizon $T>0$ and every stopping parameter $M>0$.
Suppose Assumption~\ref{ass:strong solution} - \ref{ass:Polycube} hold.
For every initial error satisfying $\EE[\|e_0^*-\hat{e}_0\|]<\varepsilon$ with $0<\varepsilon\le 1$, there exists a constant $c>0$, a stopping time $0<\tau_M\le T$ a.s.\ satisfying $0<\mathbb{E}[\tau_M]\leq O( \min\{T,M^q\})$, and a  $\rho$-MLPs $\hat\alpha$ such that the processes $\hat{E}_t$ in~\eqref{eq:SDE_X__approx} with $\hat\alpha$ and intial condition $\hat{E}_0 = \hat{e}_0$ satisfies 
\begin{align}
\label{eq:thrm:Aprx}
    \mathbb{P}\biggl(
            \sup_{0\le t \le \tau}
                \|E_t^*-\hat{E}_t\|
                +
                \|\alpha_t^{\star}-\hat{\alpha}_t\|
            \le 
                2\sqrt{\varepsilon}
        \biggr)
    & \ge 
        1
        - 
        c \sqrt{\varepsilon}
.
\end{align}
Similarly, $\tau$ can be chosen to be ``small enough'', i.e.\ $\mathbb{E}[\tau]\leq O(\varepsilon^{q(1-s/{2(1+d_E + d)})})$, so that $\hat{F}$ need not have more that $\tilde{\mathcal{O}}(1/\varepsilon)$ non-zero (trainable) parameters.
\end{theorem}

\subsection{Proofs}\label{ssec: proof of lemmas}

\begin{proof}[{Proof of Lemma~\ref{lem:Universality_Neural_SDE}}]
We show the proof in 6 steps.

\paragraph{Step 1 - Setup}
Fix $M>0$, to be set retroactively.
At time $t$, consider the ``parallelized'' map
\begin{align*}
F:(t, e^*,b) &\mapsto \big(\mu^E(t,
e^*,\alpha^*(t,e^*,b),b
),\sigma^E(t,e^*,\alpha^*(t,e^*,b),b
),\alpha^*(t,e^*,b)\big).
\end{align*}
In other words, we view the approximation problem as if we directly approximate the functional $F$, which itself contains the (optimal) control $\alpha$. Thus, we aggregate the estimation for $\alpha$ together, to avoid the approximation error of compositing with an approximate control, namely our approximation $\hat{\alpha}$ of $\alpha$.

From Assumption~\ref{ass:strong solution}, recall that $\mu^E$ and $\sigma^E$ are both $s$-order smooth. Therefore, within the rectangle region with dimension $d^*=1+d_E+d$ 
xcentered at the origin with length $2M$, by  Arz\'{e}la-Ascoli Theorem, the optimizer $\alpha^*$ is also $s$-order smooth. 

Let $\hat\alpha$ be the  $\rho$-MLP as inour setup  with the same domain as and $\alpha^*$, to be fixed retroactively depending on $M$. With $\hat{F}$ defined with respect to $\alpha$, let $\hat{E}$ be the approximated controlled SDE by~\eqref{eq:SDE_X__approx}.  We consider the stopping time $\tau_M$ by
\begin{align}
\label{eq:stopping_time}
{\tau}_M&= \tau_M^E\wedge\tau_M^\alpha\wedge\tau^B_M\wedge T,
\\
\mbox{where}\quad
\tau_M^E &= \min\{ \inf\{t>0:\, E_t^*,\hat{E}_t \not\in [-M,M]^{d_E}\}, \notag\\
\tau_M^\alpha&=\inf\{t>0:\, \alpha_t^*,\hat{\alpha}_t \not\in [-M,M]^{d_\alpha}\}, \notag\\
\tau_M^B&=\inf\{t>0:\, B_t \not\in [-M,M]^{d_\alpha}\}.
\end{align}
Note that, we choose $M$ such that $\|e^*_0\|<M$, i.e. we have a relatively nice system based on the stable formulation Section~\ref{ssec:generator}, then $\tau_M>0$ with probability 1. 

\paragraph{Step 2 - Approximation of $F$}
Let $D$ be the integer that for each $(t, e,b)$, $F(t,e,b)$ is a $D$-dim vector. 
By~\cite[Lemma 5.3]{petersen2024mathematical}, for every $K>0$, there exists an MLP $\hat{F}^{\operatorname{ReLU}}$ which is the same as $\hat{F}$ but the activation function replaced with $\operatorname{ReLU}$,  and depth $C_1K\log(K)$ and width $2DC_2 K\log(K)$, such that
\begin{equation*}
\sup_{(t,e,b)\in\mathcal{D} }\,\|F(t, e,b)-\hat{F}^{ReLU}(t,e,b)\| \leq \frac{2DMC_3}{K^{s/d^*}}.
\end{equation*}
Here $C_1$,$C_2$ and $C_3$ only depend on dimension parameters $s$ and $d^*$ by~\cite[Theorem 1.1]{lu2021deep}. 

Next, since $\rho$ is a non-degenerate sigmoidal activation function then by~\cite[Theorem 1]{zhang2024deep} there exists an MLP $\hat{F}$ whose width is at most three times the width and at most twice the depth of $\hat{F}^{\operatorname{ReLU}}$ satisfying
$$
\sup_{(t,e,b)\in\mathcal{D}}\, \|\hat{F}(t,e,b)-\hat{F}^{\operatorname{ReLU}}(t,e,b)\|\leq \frac{2DMC_3}{K^{s/d^*}}.
$$
Finally, 
we deduce the bound
\begin{align}\label{eq:approximation_bound__componentwise}
\sup_{(t,e,b)\in\mathcal{D}}\|\alpha^*(t,e,b)-\hat{\alpha}(t,e,b)\| \leq\sup_{(t,e,b)\in\mathcal{D}}\|F(t,e,b)-\hat{F}(t,e,b)\| \leq \frac{4DMC_3}{K^{s/d^*}}.
\end{align}
We denote the approximation error 
$\varepsilon=4DMC_3/K^{s/d^*}$, and notice that $\varepsilon$ scales with the process region parameter $M$, and also let $L_{\alpha^*}$ denote the Lipshitz constant of $\alpha^*$ on domain $\mathcal{D}=[0,T]\times [-M,M]^{d_E}\times [-M,M]^d$.

\paragraph{Step 3 - Approximation of Control}

Notice that
\begin{align*}
&\qquad\sup_{0\leq t\leq \tau_M}\|\alpha^*_t - \hat\alpha_t\|
\\&= \sup_{0\leq t\leq \tau_M} \|\alpha^*(t,E^*_t, B_t) - \hat\alpha(t, \hat{E}_t, B_t)\|
\\& \leq \sup_{0\leq t\leq \tau_M} \|\alpha^*(t,E^*_t, B_t) - \alpha^*(t, \hat{E}_t, B_t)\|
+\sup_{0\leq t\leq \tau_M}\|\alpha^*(t,\hat{E}_t,B_t) - \hat\alpha(t,\hat{E}_t,B_t)\|
\\&\leq  L_{\alpha^*}\sup_{0\leq t\leq \tau_M}\|E_t^* - \hat{E}_t\| +\sup_{(t,e,b)\in\mathcal{D}}\|\alpha^*(t,e,b) - \hat\alpha(t, e,b)\|
\\& \leq L_{\alpha^*}\sup_{0\leq t\leq \tau_M}\|E_t^* - \hat{E}_t\| +\epsilon.
\end{align*}
Thus we need to consider the estimation of the expected bound $\EE[\sup_{0\leq t\leq \tau_M}\|E_t^* - \hat{E}_t\| ]$ in the following step. 

\paragraph{Step 4 - Expectation Bound}
Consider the randomly stopped processes $E_{t\wedge\tau_M}$ and $\hat{E}_{t\wedge\tau_M}$, then with the correct initial information $\hat{e}_0=e^*_0$,
\begin{align*}
{E}^*_{t\wedge\tau_M}-\hat{E}_{t\wedge\tau_M}
&=\int_0^{t\wedge\tau_M} \left[\mu^E(u,E_u^*,\alpha_u^*,B_u) - \mu^E(u,\hat{E}_u,\hat{\alpha}_u,B_u) \right] du 
\\&\quad+\int_0^{t\wedge\tau_M}  
\left[\sigma^E(u,E_u^*,\alpha_u^*,B_u) - \sigma^E(u,\hat{E}_u,\hat{\alpha}_u,B_u) \right] dB_u    
\end{align*}
We first control $\sup_{0\le t\le \tau_M}\Big\|\int_0^{t}  
\left[\sigma^E(u,E_u^*,\alpha_u^*,B_u) - \sigma^E(u,\hat{E}_u,\hat{\alpha}_u,B_u) \right] dB_u \Big\|$.
By the Burkholder-Davis-Gundy (BDG) inequality with stopping times and a constant $M_2$ only depends on the dimension $d_E$, we can obtain that
\begin{align*}
&\qquad\mathbb{E}\Big[
\sup_{0\le t\le \tau_M}\Big\|\int_0^t \left[\sigma^E(E_u^*,\alpha_u^*,B_u) - \hat\sigma(\hat{E}_u,\hat{\alpha}_u,B_u)\right] dB_u\Big\|\Big]\\
& \leq M_2 \mathbb{E}\Big[\Big(\int_0^{\tau_M}\Big\|\sigma^E(u,E_u^*,\alpha_u^*,B_u) - \sigma^E(u,\hat{E}_u,\hat{\alpha}_u,B_u)\Big\|^2 du\Big)^{1/2}\Big]
\\&\leq
M_2L_{\sigma^E}\left[\mathbb{E}\Big[\Big(\int_0^{\tau_M}\Big\|\hat{E}_u-E^*_u\Big\|^2du \Big)^{1/2}\Big]
+\mathbb{E}\Big[\Big(\int_0^{\tau_M}\Big\|\hat{\alpha}_u-\alpha^*_u\Big\|^2du \Big)^{1/2}\Big]
\right]
\\
&\leq M_2L_{\sigma^E}\sqrt{T}(1+L_{\alpha^*})\mathbb{E}\left[\sup_{0\leq t\leq \tau_M}\|E^*_t- \hat{E}_t \|\right] \\&\qquad+M_2L_{\sigma^E}(1+L_{\alpha^*})\EE\left(\int_0^{\tau_M}\sup_{(t,e,b)\in\mathcal{D}}\|\alpha^*(t,e,b) - \hat\alpha(t,e,b)\|^2 dt\right)^{1/2}
\\&\leq \sqrt{T} M_2L_{\sigma^E}\left[(1+L_{\alpha^*})\mathbb{E}\left[\sup_{0\leq t\leq \tau_M}\|E^*_t- \hat{E}_t \|\right] +\EE\left[\sqrt{\frac{\tau_M}{T}}\right]\varepsilon \right].
\end{align*}
Similarly, 
\begin{align*}
&\quad \EE\left[\sup_{0\leq t\leq \tau_M} \int_0^t \left\|\mu^E(u,E_u^*,\alpha_u^*,B_u) - \mu^E(u,\hat{E}_u,\hat{\alpha}_u,B_u)\right\| du\right]
\\&\leq 
L_{\mu^E}T(1+L_{\alpha^*})\EE[\sup_{0\leq t\leq\tau_M}\|E^*_t-\hat{E}_t\|] 
\\&\qquad+L_{\mu^E} \EE\left[\int_0^{\tau_M}  \sup_{(t,e,b)\in\mathcal{D}}\|\alpha^*(t,e,b) - \hat\alpha(t,e,b)\| dt\right]
\\&\leq T L_{\mu^E}\left[(1+L_{\alpha^*})\EE[\sup_{0\leq t\leq\tau_M}\|E^*_t-\hat{E}_t\|] + \frac{\EE[\tau_M]}{T}\varepsilon\right]
\\&\leq T L_{\mu^E}\left[(1+L_{\alpha^*})\EE[\sup_{0\leq t\leq\tau_M}\|E^*_t-\hat{E}_t\|] + \EE\left[\sqrt{\frac{\tau_M}{T}}\right]\varepsilon\right],
\end{align*}
since $0<\tau_M/T\leq 1$ almost surely.
Consequently, 
\begin{align}
\label{eq:mart_part_controled__preYoga}
&\qquad\EE[\sup_{0\leq t\leq\tau_M}\|E^*_t-\hat{E}_t\|]
\\&\leq \EE\left[\sup_{0\leq t\leq \tau_M} \int_0^t \left\|\mu^E(u,E_u^*,\alpha_u^*,B_u) - \mu^E(u,\hat{E}_u,\hat{\alpha}_u,B_u)\right\| du\right]\notag\\
&\qquad
+\mathbb{E}\Big[
\sup_{0\le t\le \tau_M}\Big\|\int_0^t \left[\sigma^E(E_u^*,\alpha_u^*,B_u) - \hat\sigma(\hat{E}_u,\hat{\alpha}_u,B_u)\right] dB_u\Big\|\Big]\notag\\
&\leq (\sqrt{T} M_2L_{\sigma^E}+TL_{\mu^E})
\left[(1+L_{\alpha^*})\EE[\sup_{0\leq t\leq\tau_M}\|E^*_t-\hat{E}_t\|]
+\EE\left[\sqrt{\frac{\tau_M}{T}}\right]\varepsilon\right].
\end{align}
Therefore, for $T>0$ being relatively small enough such that 
$$
1-(\sqrt{T} M_2L_{\sigma^E}+TL_{\mu^E})
(1+L_{\alpha^*}) >0,
$$
~\eqref{eq:mart_part_controled__preYoga} implies that 
\begin{align}
\label{eq:pre_recombine}    
\EE\left[\sup_{0\leq t\leq\tau_M}\|E^*_t-\hat{E}_t\|\right]
&\leq \varepsilon\frac{\EE\left[\sqrt{\frac{\tau_M}{T}}\right](\sqrt{T} M_2L_{\sigma^E}+TL_{\mu^E})}{1-(\sqrt{T} M_2L_{\sigma^E}+TL_{\mu^E})(1+L_{\alpha^*})}
\notag\\&\leq \frac{\varepsilon M^{q/2} c_+(M_2L_{\sigma^E}+\sqrt{T}L_{\mu^E})} {1-(\sqrt{T} M_2L_{\sigma^E}+TL_{\mu^E})(1+L_{\alpha^*})},
\end{align}
where we have used the bound for the $\EE[\tau_M]\leq\EE[\tau_M^E] \leq c_+M^q$.

\paragraph{Step 5 - High Probability Guarantees}
By Markov inequality, we can obtain that
\begin{align*}
&\qquad\mathbb{P}\left[\sup_{0\leq t\leq\tau_M}\left[\|E^*_t-\hat{E}_t\|+\|\alpha^*_t - \hat\alpha_t\|\right]\geq 2\epsilon\right]
\\&\leq\mathbb{P}\left[\sup_{0\leq t\leq\tau_M}\left[\|E^*_t-\hat{E}_t\|+\|\alpha^*(t, E^*_t,B_t)- \alpha^*(t,\hat{E}_t,B_t)\|\right]\geq \epsilon\right]
\\&\leq \mathbb{P}\left[\sup_{0\leq t\leq\tau_M}\left[\|E^*_t-\hat{E}_t\|\right]\geq \frac{\varepsilon}{1+L_{\alpha^*}}\right]
\\&\leq \frac{1+L_{alpha^*}}{\varepsilon}\EE\left[\sup_{0\leq t\leq\tau_M}\right]\|E^*_t-\hat{E}_t\|
\\&\leq\frac{ M^{q/2} c_+(1+L_{\alpha^*})(M_2L_{\sigma^E}+\sqrt{T}L_{\mu^E})} {1-(\sqrt{T} M_2L_{\sigma^E}+TL_{\mu^E})(1+L_{\alpha^*})}.
\end{align*}
Therefore, for every small $\delta>0$, we can choose $M_\delta>0$ being small enough such that 
$$
\frac{ (M_\delta)^{q/2} c_+(1+L_{\alpha^*})(M_2L_{\sigma^E}+\sqrt{T}L_{\mu^E})} {1-(\sqrt{T} M_2L_{\sigma^E}+TL_{\mu^E})(1+L_{\alpha^*})}<\delta.
$$
Then we obtain the concentration-type inequality 
\begin{align*}
\mathbb{P}\left[\sup_{0\leq t\leq\tau_{M_\delta}}\left[\|E^*_t-\hat{E}_t\|+\|\alpha^*_t-\hat\alpha_t\|\right]< 2\epsilon\right]>1-\delta.
\end{align*}

\paragraph{Step 6 - Light MLPs on Small Time Horizons}
Lastly, we want to show the number of parameters in our approximation $\hat\alpha$, i.e. the number of non-zero parameters in the $\rho$-MLP, is small. To wit, retroactively setting $
M= \min\{M_\delta,\varepsilon^{1-s/(2(1+d_E+d))}\}
$ and we can obtain that
\begin{align}
\label{eq:epsilon_delta_done}
\mathbb{P}\left[\sup_{0\leq t\leq\tau_M}\left[\|E^*_t-\hat{E}_t\|+\|\alpha^*_t-\hat\alpha_t\|\right]< 2\epsilon\right]>1-\delta.
\end{align}
This choice of $M$ implies that together with $\varepsilon= 4DM{C}_3/K^{s/d^*}$ and the fact that $\hat{F}$ has depth $C_1K\log(K)$, and width $4C_2 K\log(K)$ imply that $\hat{F}$ has depth and width $\mathcal{O}\big(
\varepsilon^{-1/2}\log(\varepsilon)
\big)$; whence $\hat{F}$ must have at-most $\mathcal{O}\big(\varepsilon^{-1}\log(\varepsilon)^2\big)=\tilde{\mathcal{O}}(\varepsilon^{-1})$ non-zero (trainable) parameters.
\end{proof}

\begin{proof}[{Proof of Lemma~\ref{lem:Simple_Pertrubation_Lemma}}]
By the form of the perturbation bound of \cite[Theorem 10.6.4]{KuoStochasticIntegration} of~\cite[and Remark 10.6.5]{KuoStochasticIntegration} obtained using Doob's sub-martingale inequality, we have that 
\begin{equation}
\label{eq:sub_pertrubation__bound}
\mathbb{E}\biggl[\sup_{0\le t\le T} \|E_t-\tilde{E}_t\|^2\biggr]\leq \tilde{c}_{T,\mu^E,\sigma^E}\EE[\|e_0-\tilde{e}_0\|^2]
\end{equation}
where we choose $\tilde{c}_{T,\mu^E,\sigma^E} =   3 e^{\max\{L_{\mu^E},L_{\sigma^E}\}^2 (4+T)T}>0$.  Since we have assumed that $\EE[\|e_0-\tilde{e}_0\|^2]\le \varepsilon^2$, then~\eqref{eq:sub_pertrubation__bound} implies that 
\begin{equation*}
\mathbb{E}\biggl[\sup_{0\le t\le T}\,\|E_t-\tilde{E}_t\|\biggr]\leq \mathbb{E}\biggl[\sup_{0\le t\le T}\,\|E_t-\tilde{E}_t\|^2\biggr]^{1/2}
\leq c_{T,\mu^E,\sigma^E}\varepsilon
\end{equation*}
where $c_{T,\mu^E,\sigma^E}= \sqrt{\tilde{c}_{T,\mu^E,\sigma^E}}>0$.  Markov's inequality yields that for every $\zeta>0$,
\begin{equation*}
\mathbb{P}\biggl[\sup_{0\le t\le T}\,\|E_t-\tilde{E}_t\|\leq\zeta\biggr]\geq  1-\frac{c_{T,\mu^E,\sigma^E}\varepsilon}{\zeta},
\end{equation*}
which is our conclusion.
\end{proof}

\begin{proof}[{Proof of Theorem~\ref{thrm:Aprx}}]

Notice that $\EE[\|e_0^* - \hat{e}_0]<\epsilon$. Let $\tilde{E}$ denote the solution to the SDE with the correct initial condition $\tilde{e}_0 = e^*_0$ and the approximated control $\hat\alpha$. 
By Lemma~\ref{lem:Simple_Pertrubation_Lemma} we have that: for every $\zeta,T>0$ and $0<\varepsilon<1$
\begin{equation}
\label{eq:perturbation_control}
\mathbb{P}\biggl[\sup_{0\leq t\leq T}\,\|E_t^*-\tilde{E}_t\|\leq\zeta\biggr]
\geq 1-\frac{c_{T,\mu,\sigma}\varepsilon}{\zeta}.
\end{equation}
Then for every failure probability $\tilde{\delta}>0$, there exists a stopping time $\tau$ satisfying $0<\mathbb{E}[\tau]\le \min\{T,M^q\}$, and a $\rho$-MLPs $\hat\alpha$ such that the processes $\hat{E}$ with $\hat\alpha$ as in~\eqref{eq:SDE_X__approx} and~\eqref{eq:SDE_X__controlledprocess}, satisfy
\begin{equation}
\label{eq:approx_SDE}
\mathbb{P}\left[\sup_{0\leq t\leq\tau_M}\left[\|E^*_t-\tilde{E}_t\|+\|\alpha^*_t-\hat\alpha_t\|\right]< 2\epsilon\right]\geq1- \tilde\delta.
\end{equation}
Taking a union bound over~\eqref{eq:perturbation_control} and~\eqref{eq:approx_SDE} we find that
\begin{align*}
\mathbb{P}\biggl[\sup_{0\le t \le \tau_M}\left[\|E^*_t-\hat{E}_t\|+\|\alpha_t^*-\hat{\alpha}_t\|\right]\leq 2\varepsilon + \zeta\biggr]\geq 1-\delta-\frac{c_{T,\mu^E,\sigma^E}\varepsilon}{\zeta}.
\end{align*}
Retroactively setting $\zeta= \sqrt{\varepsilon}=\delta$ we find that
\begin{align*}
\mathbb{P}\biggl[\sup_{0\le t \le \tau_M}\left[\|E^*_t-\hat{E}_t\|+\|\alpha_t^*-\hat{\alpha}_t\|\right]\leq 2\varepsilon + \sqrt{\epsilon}\biggr]\geq
1-(1+{c_{T,\mu^E,\sigma^E}})\sqrt{\varepsilon}
.
\end{align*}
Now, since $\varepsilon\in (0,1]$ then $\sqrt{\varepsilon}\ge \varepsilon$, whence~\eqref{eq:union_bound_power} implies the cleaner bound
\begin{align}
\label{eq:union_bound_power}
\mathbb{P}\biggl[\sup_{0\le t \le \tau_M}\left[\|E^*_t-\hat{E}_t\|+\|\alpha_t^*-\hat{\alpha}_t\|\right]\leq 3\sqrt{\epsilon}\biggr]\geq
1-(1+{c_{T,\mu^E,\sigma^E}})\sqrt{\varepsilon}
\end{align}
Retroactively setting, $\delta = \sqrt{\varepsilon}$ yields the first conclusion.  
The second conclusion, on the smallness of $\hat{F}$ given an appropriate choice of $\tau$, is implied by the second conclusion of Lemma~\ref{lem:Universality_Neural_SDE}.  This concludes our proof.
\end{proof}

\begin{proof}[Proof of Theorem~\ref{thrm:AprxBaby}]
We can directly notice that Theorem~\ref{thrm:AprxBaby} follows as a corollary as Theorem~\ref{thrm:Aprx}
\end{proof}
\section{Additional Implementation Details}
\label{a:AlgosExtras}

This appendix contains additional details on the implementations and streamlined version of our algorithms in special cases.  
These models in our experiments were trained using the Virtual Machine on Google Cloud Platform with 6 CPUs and 24 GB memory. The codes containing the choice of the neural network architectures, the settings of hyperparameters, the initialization of network parameters, and all other implementation details can be found here: 
\href{https://github.com/xf-shi/Reinforced-GAN}{https://github.com/xf-shi/Reinforced-GAN}.


\begin{algorithm}[H]
\caption{Update Dynamics of Generator for LQ preference~\label{algo:generator for power cost}}
\begin{algorithmic}
\STATE{\textbf{Input: } update rule for $(\mu_{t_k}, \sigma_{t_k})=F^{k}(X_{1,t_k},\ldots, X_{N,t_k}, B_{t_k})$;}
\STATE{\hspace{38pt} parametrization: $\dot\varphi_{n,t_k}^{\theta_{n,k}^{\texttt{gen}}}=F^{\theta_{n,k}^{\texttt{gen}}}(X_{t_k}^{\theta_{n}^{\texttt{gen}}},B_{t_k}), n\in\mfN$;}
\STATE{\hspace{38pt} initial value for adjoint backward component $Y_{n,t_0} = y_{n,0}$;}
\STATE{\hspace{38pt} sample path $\Delta B$ with size $\texttt{batch\_size}\times (K+1)\times d$ ;}
\STATE{$\varphi_{n,t_0}=\bar\gamma s/\gamma_n, X_{n,t_0} = (0, W_{n,0}), J_n(\dot\varphi_n) = 0$ for each $n\in\mfN$}
\STATE{$B_{t_0} = 0, k=0$;}
\WHILE{$k\leq K$}
\STATE{for each $n\in\mfN$ in parallel: }
\STATE{\hspace{8pt} update $\xi_{n,t_k} = \xi_n B_{t_k}, (\mu_{n,t_k}, \sigma_{n,t_k}) = F^{k}(X_{1,t_k},\ldots, X_{N,t_k}, B_{t_k})$;}
\STATE{\hspace{8pt} $\dot\varphi_{n,t_k}^{\theta_{n,k}^{\texttt{gen}}}=F^{\theta_{n,k}^{\texttt{gen}}}(X_{n,t_k}^{\theta_{n}^{\texttt{gen}}},B_{t_k})$;}
\STATE{\hspace{8pt} $J_n +=\mu_{t_k}\varphi_{t_k}^{\theta_{n}^{\texttt{gen}}} -\frac{\gamma_n}{2}\left(\sigma_{t_k}\varphi_{n,t_k}^{\theta_{n}^{\texttt{gen}}}+\xi_{n,t_k}\right)^2-\frac{\lambda}{q}\left|\dot\varphi_{n,t_k}^{\theta_{n,k}^{\texttt{gen}}}\right|^q$;}
\STATE{\hspace{8pt} $\varphi_{n,t_{k+1}}^{\theta_{n,k+1}^{\texttt{gen}}}= \varphi_{n,t_k}^{\theta_{n,k}^{\texttt{gen}}}+\dot\varphi_{n,t_k}^{\theta_{n,k}^{\texttt{gen}}}\Delta t$;}
\STATE{\hspace{8pt} $X_{n,t_{k+1}}^{\theta_{n}^{\texttt{gen}}} = \varphi_{n,t_{k+1}}^{\theta_{n,k+1}^{\texttt{gen}}}-\frac{\bar\gamma}{\gamma_n} s+\frac{\xi_n}{\alpha}B_{t_k}$;}
\STATE{$B_{t_{k+1}} = B_{t_k} + \Delta B_{t_k}$;}
\STATE{$k++$;}
\ENDWHILE
\STATE{$\texttt{Loss}_{\texttt{gen}}(\theta^{\texttt{gen}}) =- \sum_{n\in\mfN} J_n/\texttt{batch\_size}$;}
\STATE{\textbf{Output: } $\texttt{Loss}_{\texttt{gen}}(\theta^{\texttt{gen}})$ with gradient information.}
\end{algorithmic}
\end{algorithm}

\begin{algorithm}[H]
\caption{Update Dynamics of Discriminator~\label{algo:discriminator for power cost}}
\begin{algorithmic}
\STATE{\textbf{Input: } update rule for  Agent-$n$'s $\dot\varphi_{n,t_k}=F^{n,k}(X_{t_k},B_{t_k})$;}
\STATE{\hspace{38pt} parametrization: $(\mu^{\theta_k^{\texttt{dis}}}_{t_k}, \sigma^{\theta_k^{\texttt{dis}}}_{t_k})=F^{^{\theta_k^{\texttt{dis}}}}(X_{1,t_k},\ldots, X_{N,t_k}, B_{t_k})$;}
\STATE{\hspace{38pt} initial value for stock price $S_{t_0}^{\theta^{\texttt{dis}}} = S_{0}$;}
\STATE{\hspace{38pt} sample path $\Delta B$ with size $\texttt{batch\_size}\times (K+1)\times d$ ;}
\STATE{$\varphi_{n,0}=\bar\gamma s/\gamma_n, X_{n,t_0} = 0, B_{t_0} = 0, J_n(\dot\varphi_n) = 0, k=0$;}
\STATE{\# Forward pass for forward state variable $X_n, n\in\mfN$: }
\STATE{$X_{n,t_0} = (0, W_{n,0})$ for each $n\in\mfN$}
\WHILE{$k\leq K$}
\STATE{\textbf{if} \emph{expression of $\mu_t$ is known} \textbf{then}}
\STATE{\hspace{8pt} $(\_\_\_\_,\sigma^{\theta_k^{\texttt{dis}}}_{t_k})=F^{^{\theta_k^{\texttt{dis}}}}(X_{1,t_k},\ldots, X_{N,t_k}, B_{t_k})$;}
\STATE{\hspace{8pt} update $\mu^{\theta_k^{\texttt{dis}}}_{t_k}$ via~\eqref{eq: mu given} and group $(\mu^{\theta_k^{\texttt{dis}}}_{t_k}, \sigma^{\theta_k^{\texttt{dis}}}_{t_k})$;}
\STATE{\textbf{else}}
\STATE{\hspace{8pt} $(\mu^{\theta_k^{\texttt{dis}}}_{t_k}, \sigma^{\theta_k^{\texttt{dis}}}_{t_k})=F^{^{\theta_k^{\texttt{dis}}}}(X_{1,t_k},\ldots, X_{N,t_k}, B_{t_k})$;}
\STATE{\textbf{end}}
\STATE{$S_{t_{k+1}}^{\theta^{\texttt{dis}}} = S_{t_k}^{\theta^{\texttt{dis}}} + \mu^{\theta_k^{\texttt{dis}}}_{t_k} \Delta t+\sigma^{\theta_k^{\texttt{dis}}}_{t_k}\Delta B_{t_k}$;}
\STATE{for each $n\in\mfN$ in parallel: }
\STATE{\hspace{8pt} $\dot\varphi_{n,t_k}=F^{n,k}(X_{n,t_k},B_{t_k})$;}
\STATE{\hspace{8pt} $\varphi_{n,t_{k+1}}= \varphi_{n,t_k}+\dot\varphi_{n,t_k}\Delta t$; }
\STATE{\hspace{8pt} $X_{n,t_{k+1}} = \varphi_{n,t_{k+1}}-\frac{\bar\gamma}{\gamma_n} s+\frac{\xi_n}{\alpha}B_{t_k}$;}
\STATE{$B_{t_{k+1}} = B_{t_k} + \Delta B_{t_k}$;}
\STATE{$k++$;}
\ENDWHILE
\STATE{\# Backward pass for adjoint backward adjoint component $Y_n, n\in\mfN$:}
\STATE{$k=K$;}
\STATE{$Y_{n,t_K} = 0$ for each $n\in\mfN$;}
\WHILE{$k\geq0$}
\STATE{for each $n\in\mfN$ in parallel: }
\STATE{\hspace{8pt} $I_{n,t_k} = \sign\left(Y_{n,t_k}\right)\left|\frac{Y_{n,t_k}}{\lambda}\right|^{\frac{1}{q-1}}$;}
\STATE{\hspace{8pt} $Y_{n,t_{k-1}} = Y_{n,t_k}+\left(\mu^{\theta_k^{\texttt{dis}}}_{t_k}-\gamma_n \sigma^{\theta_k^{\texttt{dis}}}_{t_k} (\varphi_{n,t_k} \sigma^{\theta_k^{\texttt{dis}}}_{t_k} +\xi_n B_{t_k})\right)\Delta t$;}
\STATE{$k--$;}
\ENDWHILE
\STATE{$\texttt{Loss}_{\texttt{dis}}(\theta^{\texttt{dis}}) = \left[\|S_{t_K}^{\theta^{\texttt{dis}}} -\alpha B_{t_K} - \beta T \|^2+\sum_{k=0}^K \|\sum_{n\in\mfN}I_{n,t_k}\|^2\right]/\texttt{batch\_size}$}
\STATE{\textbf{Output: } $\texttt{Loss}_{\texttt{dis}}(\theta^{\texttt{dis}})$ with gradient information.}
\end{algorithmic}
\end{algorithm}


\bibliographystyle{siamplain}
\bibliography{referencesGAN}
\end{document}